\documentclass[11pt,a4paper]{article}
\usepackage{graphicx}
\usepackage{amssymb}
\usepackage{amsmath}
\usepackage{amsfonts}
\usepackage{dsfont}
\usepackage{mathtools}
\usepackage{array}
\usepackage{rotating}
\usepackage{bbold,amsfonts}
\allowdisplaybreaks

\usepackage[utf8]{inputenc}
\usepackage{bm}
\usepackage{xcolor}
\usepackage{float}
\usepackage{braket}
\usepackage{placeins}
\usepackage[height=8.8in,width=6.45in]{geometry}
\usepackage[font=small,labelfont=bf]{caption}
\usepackage[hidelinks]{hyperref}
\usepackage{booktabs,float,slashed}
\usepackage{standalone}
\usepackage{caption}
\usepackage{subcaption}
\usepackage{makecell}
\usepackage[maxbibnames=99,sorting=none,giveninits=true,backend=biber,style=numeric-comp,sortcites,doi=false,hyperref=true]{biblatex}
\renewbibmacro{in:}{}
\usepackage{hyperref}
\DeclareFieldFormat{doilink}{\iffieldundef{doi}{#1}{\href{https://doi.org/\thefield{doi}}{#1}}}
\DeclareFieldFormat[article,periodical]{volume}{\mkbibbold{#1}}
\DeclareFieldFormat[article,periodical]{journaltitle}{#1}
\DeclareFieldFormat[article,periodical]{pages}{#1}

\usepackage{xpatch}
\xpatchbibdriver{article}
  {\usebibmacro{journal+issuetitle}%
   \newunit
   \usebibmacro{byeditor+others}%
   \newunit
   \usebibmacro{note+pages}}
  {\printtext[doilink]{%
     \usebibmacro{journal+issuetitle}%
     \newunit
     \usebibmacro{byeditor+others}%
     \newunit
     \usebibmacro{note+pages}}}
  {}{}
\numberwithin{equation}{section}
\newcommand{\vev}[1]{{\left\langle #1 \right\rangle}}

\newcommand{\beq}{\begin{equation}}
\newcommand{\eeq}{\end{equation}}

\newcommand{\overbar}[1]{\mkern 1.5mu\overline{\mkern-1.5mu#1\mkern-1.5mu}\mkern 1.5mu}

\newcommand{\vvev}[1]{\big\langle\hspace{-3pt}\big\langle #1 \big\rangle\hspace{-3pt}\big\rangle}

\DeclareMathOperator{\Tr}{Tr}
\DeclareMathOperator{\tr}{tr}

\newcommand{\bone}{\mathbf{1}}
\newcommand{\ii}{\mathrm{i}}

\makeatletter
\newcommand*{\letterdef@}{}
\newcommand*{\letterdef}[3]{%
	\def\letterdef@##1{\expandafter\newcommand\csname #1\endcsname{#2{##1}}}%
	\@tfor\@tempa :=#3\do{\expandafter\letterdef@\expandafter{\@tempa}}}
\makeatother
\letterdef{c#1} {\mathcal}{ABCDEFGHIJKLMNOPQRSTUVWXYZ} % \cX = \mathcal{X}
\letterdef{rm#1}{\mathrm} {dDeimM} % \rmX = \mathrm{X} for X in {dDmM}

\begin{document}

\begin{titlepage}

\begin{flushright}
%\footnotesize
\small
\texttt{HU-EP-23/62}
\end{flushright}

\vspace*{10mm}
\begin{center}
{\LARGE \bf 
A matrix-model approach to integrated correlators \\[2mm]
in a $\mathcal{N}=2$ SYM theory
}

\vspace*{15mm}

{\Large M. Bill\`o${}^{\,a,b}$, M. Frau${}^{\,a,b}$, A. Lerda${}^{\,c,b}$, A. Pini${}^{\,d}$}

\vspace*{8mm}
	
${}^a$ Universit\`a di Torino, Dipartimento di Fisica,\\
			Via P. Giuria 1, I-10125 Torino, Italy
			\vskip 0.3cm
			
${}^b$   I.N.F.N. - sezione di Torino,\\
			Via P. Giuria 1, I-10125 Torino, Italy 
			\vskip 0.3cm
   
${}^c$  Universit\`a del Piemonte Orientale,\\
			Dipartimento di Scienze e Innovazione Tecnologica\\
			Viale T. Michel 11, I-15121 Alessandria, Italy
			\vskip 0.3cm
   
${}^d$ Institut f{\"u}r Physik, Humboldt-Universit{\"a}t zu Berlin,\\
IRIS Geb{\"a}ude, Zum Großen Windkanal 2, 12489 Berlin, Germany  

\vskip 0.8cm
	{\small
		E-mail:
		\texttt{billo,frau,lerda@to.infn.it;alessandro.pini@physik.hu-berlin.de}
	}
\vspace*{0.8cm}
\end{center}

\begin{abstract}

In a $\cN=2$ superconformal gauge theory with matter hypermultiplets transforming in the symmetric and anti-symmetric representations of SU($N$), we study the integrated correlators of two Coulomb-branch operators and two moment-map operators using localization. In the corresponding matrix model we identify the operator associated with the integrated insertions of moment-map operators and provide for it an exact expression valid for all values of the coupling constant in the planar limit. This allows us to study the integrated correlators at strong-coupling where we show that they behave as the 2-point functions of the Coulomb-branch operators, up to an overall constant dependent only on the conformal dimensions of the latter.
The strong-coupling relation between integrated correlators and 2-point functions turns out to be the same as in $\cN=4$ SYM at large $N$, despite the reduced amount of supersymmetry in our theory.

\end{abstract}
\end{titlepage}
\setcounter{tocdepth}{2}
\tableofcontents

\section{Introduction and outline}
\label{sec:intro}
Correlation functions of primary operators in four-dimensional superconformal theories have attracted much attention from different points of view, and many important results have been obtained over the years. One of the most powerful techniques used for these investigations is supersymmetric localization\,%
\footnote{See for example \cite{Pestun:2016zxk} and references therein.}, which reduces functional integrals over fields to finite integrals over matrices \cite{Pestun:2007rz}. Exploiting localization, it has been shown \cite{Baggio:2014sna,Baggio:2015vxa,Baggio:2016skg,Gerchkovitz:2016gxx,Billo:2017glv} that the 2- and 3-point functions of primary operators in the Coulomb branch of $\cN=2$ superconformal gauge theories can be obtained from the 2- and 3-point correlators of operators defined in an interacting matrix model whose specific features depend on the field content of the superconformal gauge theory \cite{Pestun:2007rz}. In this way the computations are drastically simplified, making it possible in several instances to obtain exact results valid for all values of the coupling constant at least in the planar limit. In the case of the $\cN=4$ Super Yang-Mills (SYM) theory, the matrix model arising from localization is purely Gaussian and the calculations become even simpler. 

In the matrix model one can obviously compute also correlators with four or more operators, but these correlators do not provide information on the 4- or higher-point correlation functions of primary operators in the superconformal gauge theory. Only for the so-called ``extremal'' correlators it possible to exploit conformal invariance combined with supersymmetry and fix the functional form of the correlation functions up to coefficients that can be captured by a matrix-model calculation \cite{Gerchkovitz:2016gxx}. 

Recently, however, it has been shown that localization techniques can be used to derive \emph{integrated} correlation functions of four primaries in $\cN=4$ SYM. The argument is as follows. First, one places $\cN=4$ SYM on a 4-sphere and deforms it with a mass parameter $m$ while preserving $\cN=2$ supersymmetry. In this way one obtains the so-called $\cN=2^*$ SYM, whose partition function $\cZ_{\cN=2^*}$ can be computed using matrix model techniques \cite{Pestun:2007rz,Russo:2013kea}. Then, one can prove that \cite{Binder:2019jwn}
\begin{align}
   \partial_{\tau_p} \partial_{\,\overbar{\tau}_p}\partial_m^2 \log\cZ_{\cN=2^*}\Big|_{m=0}&= \int\!\prod_{i=1}^4dx_i~\mu(\{x_i\})\,
    \big\langle\mathcal{O}_p(x_1) \,\overbar{\mathcal{O}}_p(x_2)\,\mathcal{J}(x_3)\,
    \mathcal{J}(x_4)\big\rangle_{\cN=4}~.
    \label{integratedppJJ}
\end{align}
Here $\tau_p$ are $\overbar{\tau}_p$ are the couplings associated to chiral and anti-chiral Coulomb-branch operators $\mathcal{O}_p(x_i)$ and $\overbar{\mathcal{O}}_p(x_i)$ which are gauge-invariant superconformal primaries of dimension $p$ constructed with the scalar fields of the vector multiplet. When $p=2$, these scalar operators are in the $\mathbf{20^\prime}$ representation of the $R$-symmetry group SU(4) of $\cN=4$ SYM inside the stress-tensor multiplet. Instead, $\mathcal{J}(x_i)$ is the moment-map operator associated to the mass deformation of the $\cN=2^*$ theory on the sphere. This is a dimension 2 operator constructed from the scalar fields of the adjoint hypermultiplet which can be written as a particular combination of $\mathbf{20^\prime}$ operators (see for example \cite{Bobev:2013cja,Binder:2019jwn} for details). Therefore, for $p=2$ the relation (\ref{integratedppJJ}) provides information on the correlation function of four primaries of dimension 2 belonging to the $\mathbf{20^\prime}$ representation, integrated with a suitable integration measure $\mu(\{x_i\})$ that is fixed by conformal invariance and supersymmetry. The explicit expression of this measure can be found in \cite{Binder:2019jwn}, but it is not needed for our purposes.

If we normalize the 2-point function of the Coulomb-branch operators in the canonical way as
\begin{align}
    \big\langle \cO_p(x_1)\,\overbar{\cO}_p(x_2)\big\rangle_{\cN=4}=\frac{\cG_p^{(0)}}{\big(4\pi^2|x_1-x_2|^2\big)^p}~,
    \label{norm2point}
\end{align}
then, using the results of \cite{Binder:2019jwn}, one can show that in the planar limit
\begin{align}
  \lim_{\lambda\to\infty} \frac{\partial_{\tau_p} \partial_{\,\overbar{\tau}_p}\partial_m^2 \log\cZ_{\cN=2^*}\Big|_{m=0}}{\cG_p^{(0)}}= \frac{p-1}{2}
  \label{ratioN4}
\end{align}
where $\lambda$ is the 't Hooft coupling. This simple result follows from the strong-coupling behavior of the integrated correlators (\ref{integratedppJJ}) given in \cite{Binder:2019jwn} adapted to our conventions, and the fact that the coefficients $\cG_p^{(0)}$ do not depend on $\lambda$ in the planar limit.

Following the initial proposal of \cite{Binder:2019jwn}, the relation (\ref{integratedppJJ}) has been investigated in a series of papers \cite{Chester:2019pvm,Chester:2019jas,Chester:2020dja,Chester:2020vyz} and also generalized to  
\begin{align}
   \partial_m^4 \log\cZ_{\cN=2^*}\Big|_{m=0}&= \int\!\prod_{i=1}^4dx_i~\hat{\mu}(\{x_i\})\, \big\langle\mathcal{J}(x_1) \,\mathcal{J}(x_2)\,\mathcal{J}(x_3)\,\mathcal{J}(x_4)\big\rangle_{\cN=4}~.
    \label{integrated4J}
\end{align}
This new relation provides further information on the 4-point correlation function of $\mathbf{20^\prime}$ primaries integrated with a new measure $\hat{\mu}(\{x_i\})$.
Many features of the integrated correlators (\ref{integratedppJJ}) and (\ref{integrated4J}) have been explored in the last few years, in particular by identifying their modular and weak-coupling  properties \cite{Dorigoni:2021bvj,Dorigoni:2021guq,Collier:2022emf,Dorigoni:2022cua,Paul:2022piq,Wen:2022oky}, by introducing general gauge groups \cite{Dorigoni:2022zcr,Dorigoni:2023ezg}, by considering operator insertions with generic \cite{Brown:2023zbr} or large conformal dimensions \cite{Paul:2023rka,Brown:2023cpz,Brown:2023why,Caetano:2023zwe}. Moreover, mixed integrated correlators involving local operators and a Wilson line in $\cN=4$ SYM have been recently considered in \cite{Pufu:2023vwo,Billo:2023ncz}.

Another interesting development concerns the study of integrated correlators in superconformal gauge theories with $\cN=2$ supersymmetry. In \cite{Chester:2022sqb,Fiol:2023cml,Behan:2023fqq} the integrated 4-point functions of moment-map operators have been studied in $\cN=2$ superconformal QCD (SQCD) using localization and matrix-model techniques. Starting from the free energy of a deformed version of SQCD with massive hypermultiplets, integrated insertions of moment-map operators have been obtained by taking derivatives with respect to the hypermultiplet mass. In this way an exact relation between the fourth mass-derivatives of the free energy of the massive SQCD and the integrated 4-point function of moment-map operators has been established, in strict analogy with (\ref{integrated4J}). 

In this paper we further elaborate on the integrated correlators of $\cN=2$ SYM theories but, differently from \cite{Chester:2022sqb,Fiol:2023cml,Behan:2023fqq}, we consider mixed correlation functions between Coulomb-branch operators and moment-map operators, like those in (\ref{integratedppJJ}). We do so in a particular $\cN=2$ superconformal gauge theory, called \textbf{E}-theory \cite{Billo:2019fbi,Billo:2022xas}, in which the hypermultiplets transform in the rank-2 symmetric and anti-symmetric representations%
\footnote{We recall that in a $\cN=2$ SU($N$) SYM theory with $N_f$ fundamental, $N_S$ symmetric and $N_A$ anti-symmetric hypermultiplets, the $\beta$-function coefficient is 
$\beta_0=2N-N_f-N_S(N+2)-N_A(N-2)$.
Therefore, with the field content of the \textbf{E}-theory ($N_f=0$, $N_S=N_A=1$) the $\beta$-function identically vanishes for all $N$.} of SU($N$).
The \textbf{E}-theory is interesting because it arises with a combination of orbifold and orientifold $\mathbb{Z}_2$-projections from a parent $\cN=4$ SYM with gauge group SU($2N$) and shares many properties with the latter while having a reduced amount of supersymmetry. In fact, all \textbf{E}-theory observables involving operators that are even (or untwisted) under the $\mathbb{Z}_2$-projections coincide with those of the parent $\cN=4$ SYM in the planar limit and differ only in the sub-leading non-planar corrections. On the contrary, for all quantities that are odd (or twisted) under the orbifold or orientifold parities, the differences with respect to $\cN=4$ SYM already appear at the planar level. On top, at strong coupling the \textbf{E}-theory admits a dual description in terms of Type II B strings on AdS$_5\times S^5/\mathbb{Z}_2$ \cite{Kachru:1998ys,Ennes:2000fu} and thus it is a relatively simple playground to test the holographic correspondence when supersymmetry is not maximal.

To obtain the integrated correlators along the lines discussed above for $\cN=4$ SYM, we place the \textbf{E}-theory on a 4-sphere and then give a mass $m$ to the symmetric and anti-symmetric hypermultiplets\,%
\footnote{As we will explain in Section~\ref{sec:Estar}, we could give different masses to the two types of hypermultiplets. However, to highlight similarities and differences with respect to the $\cN=2^*$ theory in the simplest and most direct way, here we assume that both masses are equal to $m$. Furthermore, we will see that choosing different masses induces effects that are sub-leading in the large-$N$ expansion with respect to those obtained when the masses are equal.}.
The resulting massive theory, which is still $\cN=2$ supersymmetric but obviously no longer conformal, will be called $\mathbf{E}^*$-theory, and its partition function $\mathcal{Z}_{\mathbf{E}^*}$
can be studied using supersymmetric localization. Then, we compute
\begin{align}
   \partial_{\tau_p} \partial_{\,\overbar{\tau}_p}\partial_m^2 \log\cZ_{\mathbf{E}^*}\Big|_{m=0}
   \label{derivativeZE}
\end{align}
which, in analogy with (\ref{integratedppJJ}), provides information on the mixed integrated correlator among two Coulomb-branch operators and two moment-map operators of the \textbf{E}-theory. Indeed, the action of a $\cN=2$ SYM theory with massive hypermultiplets on a 4-sphere of radius $r$, contains among others a term of the form $\frac{\ii\,m_i}{r}\!\int\!d^4x\,\sqrt{g}\,\mathcal{J}^i$, where $\cJ^i$ is the moment-map operator associated to the $i$-th hypermultiplet (for details see, for example, \cite{Binder:2019jwn,Chester:2022sqb,Fiol:2023cml,Behan:2023fqq}). In our case this term becomes proportional to $m\!\int\!d^4x\,\sqrt{g}\,\mathcal{J}$, where $\cJ$ is the moment-map operator in the symmetric plus anti-symmetric representation\,%
\footnote{In SU($N$) we have $N\otimes\overbar{N}=1\oplus\mathrm{adj}$ and $N\otimes{N}=1\oplus \mathrm{symm}\oplus$ anti-symm. Thus, the combination symmetric plus anti-symmetric is very similar, but obviously not identical, to the adjoint representation. For this reason we use the same symbol $\cJ$ for the moment-map operators in the two cases.}. Thus, derivatives with respect to $m$ correspond to integrated insertions of $\cJ$, and the integrated correlator associated to (\ref{derivativeZE}) can be schematically written as
\begin{align}
    \int\!\prod_{i=1}^4dx_i~\mu^\prime(\{x_i\})\,
    \big\langle\mathcal{O}_p(x_1) \,\overbar{\mathcal{O}}_p(x_2)\,\mathcal{J}(x_3)\,
    \mathcal{J}(x_4)\big\rangle_{\mathbf{E}}
    \label{intOOJJ}
\end{align}
where the integration measure is again fixed by conformal invariance and supersymmetry. 

The main goal of this paper is to study the quantity in (\ref{derivativeZE}). We do so by using the matrix-model approach that allows us to obtain an explicit expression for the partition function $\cZ_{\mathbf{E}^*}$ that is valid for all values of the 't Hooft coupling in the planar limit. Then, in analogy with (\ref{ratioN4}), we argue that
\begin{align}
  \lim_{\lambda\to\infty}  \frac{\partial_{\tau_p} \partial_{\,\overbar{\tau}_p}\partial_m^2 \log\cZ_{\mathbf{E}^*}\Big|_{m=0}}{\cG_p}= \frac{p-1}{2}
  \label{ratioE}
\end{align}
where $\cG_p$ is the normalization factor of the 2-point functions of Coulomb-branch operators in the $\mathbf{E}$-theory. We analytically check this relation for all even $p$, while for odd $p$ we provide compelling evidence of its validity based on numerical results. When $p$ is even the relation (\ref{ratioE}) can be demonstrated following the same steps as in $\cN=4$ SYM, since all quantities appearing in the calculation are even under the orbifold/orientifold projections that lead to the \textbf{E}-theory and thus are planar equivalent to those of $\cN=4$ SYM. Instead, for odd $p$ the relation (\ref{ratioE}) is highly non-trivial since it involves quantities that are odd under the orbifold/orientifold projections and are not planar equivalent to those in $\cN=4$ SYM. Indeed, both the numerator and the denominator in the left-hand side are complicated functions of the 't Hooft coupling, and the fact that at strong coupling these two functions coincide in the planar limit up to an overall numerical factor is quite remarkable.

The outline of the paper is as follows. In Section~\ref{sec:Estar} we review the matrix-model description of the \textbf{E} and $\mathbf{E}^*$-theories, and identify the matrix operator which represents the integrated insertion of two moment-map operators. As shown in \cite{Beccaria:2020hgy,Beccaria:2021hvt} the matrix model corresponding to the \textbf{E}-theory is characterized by an interaction action that in the planar limit can be written in terms of convolutions of Bessel functions with a particular kernel. This expression is exact in the coupling constant and can be fruitfully used to investigate the strong-coupling regime. It turns out that also the operator that accounts for the integrated insertion of the moment-map operators can be written in the matrix model in terms of convolutions of Bessel functions, but with a different kernel. Despite this difference, the same techniques used to study the matrix model of the $\mathbf{E}$-theory can be applied to study the mass corrections in the $\mathbf{E}^*$-theory. We do this in Section~\ref{sec:freeenergy}, where in particular we find an exact expression for the first mass correction of the free energy of the $\mathbf{E}^*$-theory which is valid for all values of the coupling constant in the planar limit. From it we then obtain the leading strong-coupling behavior of the free energy using techniques similar to those developed in \cite{Belitsky:2020qrm,Belitsky:2020qir} to study the asymptotic expansion of the octagon form factor in $\cN=4$ SYM at large coupling. 
In Section~\ref{sec:integrated} we study the matrix-model representation of the integrated correlators in a simple case corresponding to insertions of Coulomb-branch operators of conformal dimension 2, first in $\cN=4$ SYM where the underlying matrix model is Gaussian, and then in the \textbf{E}-theory. In both cases we obtain an explicit expression that we subsequently generalize in 
Section~\ref{sec:integratedp} to insertions of operators with arbitrary dimension $p$, proving (\ref{ratioE}).
Finally, in Section~\ref{sec:conclusions} we present our conclusions while in the appendices we collect many technical details that are useful to reproduce our results.

\section{The \texorpdfstring{$\mathbf{E}^*$}{}-theory and its matrix-model description}
\label{sec:Estar}

The $\mathbf{E}^*$-theory is a $\cN=2$ massive deformation of the superconformal $\mathbf{E}$-theory with gauge group SU($N$) in which the rank-2 symmetric and anti-symmetric hypermultiplets acquire a mass $m_S$ and $m_A$, respectively.
Using supersymmetric localization, the partition function of the $\mathbf{E}^*$-theory can be written as an integral over a Hermitian matrix $a$
as follows \cite{Pestun:2007rz}\,%
\footnote{In our conventions, $a=\displaystyle{\sum_{b=1}^{N^2-1}a^b\,T_b}$ where the normalization of the generators of SU($N$) is $\tr T_bT_c=\frac{1}{2}\delta_{bc}$. We have also inserted a normalization factor which clearly drops out in expectation values but which is useful to simplify some of the following formulas. 
Furthermore, we have set the radius of the 4-sphere to one.}
\begin{align}
    \cZ_{\mathbf{E}^*}=\Big(\frac{8\pi^2N}{\lambda}\Big)^{\frac{N^2-1}{2}}\int\!da~\rme^{-\frac{8\pi^2N}{\lambda}\tr a^2} \Big|Z_{\mathrm{1-loop}}(a,m_S,m_A)\,Z_{\mathrm{inst}}(a,m_S,m_A,\lambda)\Big|^2~.
    \label{Z}
\end{align}
Here $\lambda$ is the 't Hooft coupling and $Z_{\mathrm{1-loop}}(a,m_S,m_A)$ and $Z_{\mathrm{inst}}(a,m_S,m_A,\lambda)$ are the 1-loop and the non-perturbative instanton contributions. In the limit $N\to\infty$ with $\lambda$ fixed, we can neglect the instanton part and set
$Z_{\mathrm{inst}}(a,m_S,m_A,\lambda)=1$. The 1-loop term, instead, is not trivial and can be written compactly as
\begin{align}
    \Big|Z_{\mathrm{1-loop}}(a,m_S,m_A)\Big|^2=\frac{\displaystyle{\prod_{\mathbf{v}\in W(\mathrm{adj})}}
    H(\ii\,\mathbf{v}\cdot a)}{\displaystyle{ \prod_{\cR=S,A}\bigg[\!\prod_{~\mathbf{w}_{\cR}\in W(\cR)}} H( \ii\,\mathbf{w}_{\cR}\cdot a+\ii\,m_{\cR}) \bigg]}
\end{align}
where $\mathbf{v}$ are the weights of the adjoint representation of SU($N$), $\mathbf{w}_{\cR}$ are the weights of the representations $\mathcal{R}$ of the hypermultiplets (namely the rank-2 symmetric and anti-symmetric) and $H$ is defined as\,%
\footnote{Here we follow the conventions of \cite{Russo:2013kea} with $x\leftrightarrow \ii\,x$ and define, differently from \cite{Pestun:2007rz}, the function $H$ with the extra exponential factor. In this way the dependence on $(1+\gamma)$ will cancel in all formulas.} 
\begin{equation}
    H(x)= \prod_{n=1}^\infty \Big(1-\frac{x^2}{n^2}\Big)^n\,\rme^{\frac{x^2}{n}}\,\equiv\,\rme^{(1+\gamma)x^2}\,G(1+x)\,G(1-x)
    \label{His}
\end{equation}
with $G$ being the Barnes $G$-function and $\gamma$ the Euler-Mascheroni constant. 

Since we are interested in small deformations around the \textbf{E}-theory, we may expand the previous expressions for small values of $m_S$ and $m_A$. Doing so and exploiting the fact that $H$ is an even function, we easily obtain
\begin{align}
    \Big|Z_{\mathrm{1-loop}}(a,m_S,m_A)\Big|^2&= \Big|Z_{\mathrm{1-loop}}(a)\Big|^2\,\bigg(1-\frac{1}{2}\sum_{\cR=S,A}\sum_{~\mathbf{w}_{\cR}\in W(\mathcal{R})}\!\!m_{\cR}^2\,\partial^2\log H(\ii\,\mathbf{w}_{\cR}\cdot a)+...
    \bigg)
    \label{Z1loop}
\end{align}
where the ellipses stand for higher order terms in the mass expansion and 
\begin{align}
    \Big|Z_{\mathrm{1-loop}}(a)\Big|^2
    &=\frac{\displaystyle{\prod_{\mathbf{v}\in W(\mathrm{adj})}}
    H(\ii\,\mathbf{v}\cdot a)}{\displaystyle{ \prod_{\cR=S,A}\bigg[\!\prod_{~\mathbf{w}_{\cR}\in W(\cR)}} H(\ii\,\mathbf{w}_{\cR}\cdot a) \bigg]}
\end{align}
is the 1-loop term of the \textbf{E}-theory.
If we rewrite (\ref{Z1loop}) in exponential form as
\begin{align}
    \Big|Z_{\mathrm{1-loop}}(a,m_S,m_A)\Big|^2&=\exp\bigg[\!-\!\Big(\!\sum_{\cR=S,A}\sum_{~\mathbf{w}_{\cR}\in W(\mathcal{R})}\log H(\ii\,\mathbf{w}_{\cR}\cdot a)-\!\!\sum_{\mathbf{v}\in W(\mathrm{adj})}\!\!\log H(\ii\,\mathbf{v}\cdot a)\Big)\notag\\ &\qquad ~~~~\,-\frac{1}{2}\!\sum_{\cR=S,A}\sum_{~\mathbf{w}_{\cR}\in W(\mathcal{R})}\!m_{\cR}^2\,\partial^2\log H(\ii\,\mathbf{w}_{\cR}\cdot a)+...\bigg]~,
    \label{Z1loopexp}
\end{align}
we can view $|Z_{\mathrm{1-loop}}(a,m_S,m_A)|^2$ as an interaction action added to the Gaussian matrix model, that consists of a mass-independent term
\begin{align}
    \widetilde{S}_{0}=\sum_{\cR=S,A}\!\sum_{~\mathbf{w}_{\cR}\in W(\mathcal{R})}\log H(\ii\,\mathbf{w}_{\cR}\cdot a)-\!\!\!\sum_{\mathbf{v}\in W(\mathrm{adj})}\log H(\ii\,\mathbf{v}\cdot a)~,
    \label{Sint0}
\end{align}
and a tail of mass-corrections, the first of which being
\begin{align}
    \widetilde{S}_{2}=\frac{1}{2}\!\sum_{\cR=S,A}\!\sum_{~\mathbf{w}\in W(\mathcal{R})}\!m_{\cR}^2\,\partial^2\log H(\ii\,\mathbf{w}_{\cR}\cdot a)~.
    \label{Sint2}
\end{align}
Thus, for small masses the partition function of the $\mathbf{E}^*$-theory can be written as
\begin{align}
    \cZ_{\mathbf{E}^*}=\Big(\frac{8\pi^2N}{\lambda}\Big)^{\frac{N^2-1}{2}}\int\!da~\rme^{-\frac{8\pi^2N}{\lambda}\tr a^2} ~
    \rme^{-\widetilde{S}_{0}-
    \widetilde{S}_{2}+...}~.
    \label{Z1}
\end{align}
Note that if $\mathcal{R}$ were the adjoint representation, $\widetilde{S}_{0}$ would vanish and $\widetilde{S}_{2}$ would reduce to the first mass correction in the matrix model describing the $\cN=2^*$ theory \cite{Russo:2013kea}.
Using the explicit form of the weights \textbf{v}, $\mathbf{w}_S$ and $\mathbf{w}_A$,
we can write $\widetilde{S}_{0}$ and $\widetilde{S}_{2}$ in terms of the eigenvalues
$a_u$ ($u=1,\ldots,N)$ of the matrix $a$, finding
\begin{align}
    \widetilde{S}_{0}=\sum_{u\leq v}\log H(\ii\,a_u+\ii\,a_v)+\sum_{u<v}\log H(\ii\,a_u+\ii\,a_v)-2\sum_{u<v}\log H(\ii\,a_u-\ii\,a_v)
    \label{S0tildenew}
\end{align}
and
\begin{align}
    \widetilde{S}_{2}=\frac{1}{2}m_S^2\sum_{u\leq v}\partial^2\log H(\ii\,a_u+\ii\,a_v)+\frac{1}{2}m_A^2\sum_{u<v}\partial^2\log H(\ii\,a_u+\ii\,a_v)~.
    \label{S2tildenew}
\end{align}

From (\ref{Z1}) we can easily derive the mass-expansion of the free energy $\mathcal{F}_{\mathbf{E}^*}=-\log \mathcal{Z}_{\mathbf{E}^*}$, whose first terms are 
\begin{align}
    \cF_{\mathbf{E}^*}=\cF_{\mathbf{E}} +\big\langle \widetilde{S}_{2}\big\rangle
    +...
    \label{Fm}
\end{align}
where the notation $\langle f \rangle$ means the expectation value of $f$ in the matrix model of the $\mathbf{E}$-theory, namely
\begin{align}
    \langle f \rangle =\frac{\displaystyle{\int\!da~\rme^{-\frac{8\pi^2N}{\lambda}\tr a^2} ~
    \rme^{-\widetilde{S}_{0}}\,f}\phantom{\bigg|}}{\displaystyle{\int\!da~\rme^{-\frac{8\pi^2N}{\lambda}\tr a^2} ~
    \rme^{-\widetilde{S}_{0}}}\phantom{\bigg|}}~.
    \label{vevf}
\end{align}
In (\ref{Fm}) the term $\cF_{\mathbf{E}}\,\equiv\,\cF_{\mathbf{E}^*}\big|_{m_S=m_A=0}$ represents the free-energy of the \textbf{E}-theory. This has been thoroughly studied both at weak and at strong coupling in \cite{Beccaria:2022ypy,Beccaria:2023kbl} to which we refer for details. Here, instead we concentrate on the first mass correction $\langle \widetilde{S}_{2} \rangle$.
To compute it we first rescale the matrix $a$ according to
\begin{align}
    a\,~\to\,~\sqrt{\frac{\lambda}{8\pi^2 N}}\,a~,
    \label{rescaling}
\end{align}
so that the quadratic term in (\ref{vevf}) acquires a canonical Gaussian normalization, and then
expand in powers of $\lambda$ taking advantage of the formula
\begin{align}
    \log H(x)=-\sum_{n=1}^\infty\,\frac{\zeta_{2n+1}}{n+1}\,x^{2n+2}
    \label{Hexp}
\end{align}
where $\zeta_k$ is the Riemann $\zeta$-value $\zeta(k)$.
Proceeding in this way and rewriting (\ref{S0tildenew}) and (\ref{S2tildenew}) in terms of the traces of powers of $a$, after some algebra we find
\begin{align}
     \widetilde{S}_{0}\,~&\to\,~S_{0}=4\!\sum_{n,\ell=1}^\infty
     (-1)^{n+\ell}\,\frac{(2n+2\ell+1)!\,\zeta_{2n+2\ell+1}}{(2n+1)!\,(2\ell+1)!}\,\Big(\frac{\lambda}{8\pi^2 N}\Big)^{n+\ell+1} \tr a^{2n+1} \tr a^{2\ell+1}~,
     \label{S0}
\end{align}
and
\begin{align}
         \widetilde{S}_{2}\,~&\to\,~S_{2}= -\frac{m_S^2+m_A^2}{2}~\mathcal{M}-
         \frac{m_S^2-m_A^2}{2}~\mathcal{M}^\prime~,
     \label{S2}
\end{align}
where
\begin{align}
    \mathcal{M}&=-\sum_{n=1}^{\infty}\sum_{\ell=0}^{2n}(-1)^{n}\,\frac{(2n+1)!\,\zeta_{2n+1}}{(2n-\ell)!\,\ell!}\,\Big(\frac{\lambda}{8\pi^2 N}\Big)^{n} \tr a^{2n-\ell}\tr a^{\ell}~,
    \label{Mis}\\[2mm]
    \mathcal{M}^\prime&=-\sum_{n=1}^\infty (-1)^n\,(2n+1)\,\zeta_{2n+1}\,\Big(\frac{\lambda}{2\pi^2 N}\Big)^{n}
    \tr a^{2n}~.
    \label{Mprimeis}
\end{align}
A few remarks are in order. Firstly, we see that both $S_0$ and $\cM$ are quadratic in the traces of $a$ and, even if the coefficients are not the same, their structure is analogous. We therefore expect that the techniques developed in \cite{Beccaria:2020hgy,Beccaria:2021hvt} to study $S_0$ can be used also for $\cM$. Secondly, we observe that $\cM$ is very similar to the operator
$\cM_0$ that accounts for the first mass-correction in the $\cN=2^*$ theory. Indeed, the latter is given by (see for instance eq.
(5.9) of \cite{Billo:2023ncz})
\begin{align}
    \mathcal{M}_0=-\sum_{n=1}^{\infty}\sum_{\ell=0}^{2n}(-1)^{n+\ell}\,\frac{(2n+1)!\,\zeta_{2n+1}}{(2n-\ell)!\,\ell!}\,\Big(\frac{\lambda}{8\pi^2 N}\Big)^{n} \tr a^{2n-\ell}\tr a^{\ell}~,
    \label{M0foot}
\end{align}
which differs from $\cM$ in (\ref{Mis}) only by an additional factor of $(-1)^\ell$. The operator $\cM^\prime$, instead, is new and has no analogue in the $\cN=2^*$ theory.

The structure of $S_2$ in (\ref{S2}) suggests to introduce the following combinations
\begin{align}
    m^2=\frac{m_S^2+m_A^2}{2}\qquad \mbox{and}
    \qquad m^{\prime\,2}=\frac{m_S^2-m_A^2}{2}~.
    \label{mmprime}
\end{align}
Due to the analogy between $\cM$ and $\cM_0$, we easily realize that $m^2$ plays a role similar to that of the mass deformation in $\cN=2^*$, while the second combination clearly vanishes when the masses of the symmetric and anti-symmetric hypermultiplets are equal. With these definitions, 
the free energy (\ref{Fm}) becomes
\begin{align}
    \cF_{\mathbf{E}^*}&=\cF_{\mathbf{E}}-m^2\,\big\langle \mathcal{M}\big\rangle-m^{\prime\,2}\,\big\langle \mathcal{M}^\prime\big\rangle+...~.
    \label{Fm1}
\end{align}
Thus, up to quartic order in the masses the calculation of $\cF_{\mathbf{E}^*}$ reduces to the evaluation of the vacuum expectation values
$\langle \tr a^{k} \tr a^{\ell} \rangle$ and
$\langle \tr a^{k} \rangle$ in the matrix model of the \textbf{E}-theory. In turn, these vacuum expectation values can be written in terms of the vacuum expectation values in the Gaussian
matrix model, denoted by a subscript 0, according to
\begin{align}
    \big\langle
    \tr a^{k}\tr a^{\ell}\big\rangle&=\frac{\displaystyle{\int \!da~\rme^{-\tr a^2}\,\rme^{-S_{0}}\,\tr a^{k} \tr a^{\ell}}}{\displaystyle{\int \!da~\rme^{-\tr a^2}\,\rme^{-S_{0}}}}=\frac{\big\langle \rme^{-S_{0}}\,\tr a^{k} \tr a^{\ell}\big\rangle_0\phantom{\Big|}}{\big\langle \rme^{-S_{0}}\big\rangle_0\phantom{\Big|}}
    ~,
    \label{tklE}
\end{align}
and similarly for $\langle \tr a^{k} \rangle$.

\subsection{Mass-corrections in the large-\texorpdfstring{$N$}{} limit}
\label{subsec:mclN}
As discussed in \cite{Beccaria:2021hvt,Billo:2022fnb,Billo:2022lrv}, in the large-$N$ limit it is convenient to change basis and, instead of $\tr a^k$, use a new set of operators $\cP_k$ that are normal-ordered and orthonormal in the Gaussian model at leading order\,%
\footnote{From now on, we will understand that all equations hold at leading order for the $N\to\infty$. We will explicitly indicate the presence of the sub-leading corrections when necessary.} for $N\to\infty$, namely
\begin{align}
    \langle \cP_k\rangle_0=0~,
    \label{Pk0}
\end{align}
and
\begin{align}
    \langle \cP_k\,\cP_\ell\rangle_0=\delta_{k,\ell}~.
    \label{PkPl0}
\end{align}
The relation between the two basis can be worked out explicitly in full generality \cite{Rodriguez-Gomez:2016cem} and reads as follows
\begin{align}
    \tr a^k=\Big(\frac{N}{2}\Big)^{\frac{k}{2}}\,\sum_{\ell=0}^{\lfloor\frac{k-1}{2}\rfloor}
    \sqrt{k-2\ell} \,\binom{k}{\ell} \,\mathcal{P}_{k-2\ell}+
    \langle \tr a^k\rangle_0~.
    \label{avsP}
\end{align}
If we substitute this in (\ref{S0}), the interaction action $S_{0}$ takes a remarkably simple form
and becomes
\begin{align}
    S_{0}=-\frac{1}{2}\sum_{k,\ell=1}^\infty
    \cP_{2k+1}\,\mathsf{X}_{2k+1,2\ell+1}\,\cP_{2\ell+1}
    \label{S0P}
\end{align}
where the $\lambda$-dependent coefficients can be written in terms of a convolution of Bessel functions of the
first kind according to \cite{Beccaria:2020hgy,Beccaria:2021hvt}
\begin{align}
    \mathsf{X}_{n,m}=2\,(-1)^{\frac{n+m+2nm}{2}+1}\,\sqrt{nm}\int_0^\infty\!\frac{dt}{t}\,\frac{1}{\sinh(t/2)^2}\,J_n\Big(\frac{t\sqrt{\lambda}}{2\pi}\Big)\,J_m\Big(\frac{t\sqrt{\lambda}}{2\pi}\Big)
    \label{Xij}
\end{align}
for $n,m\geq2$.
Notice that while the initial expression (\ref{S0}) is an expansion in powers of $\lambda$ and
thus is valid at weak coupling, the expression (\ref{S0P}) is a resummation of the perturbative series and can be used for \emph{all} values of $\lambda$. Indeed, the coefficients (\ref{Xij})
are well-defined for any $\lambda$ and admit a simple asymptotic expansion at strong coupling that can be obtained using the Mellin-Barnes method \cite{Beccaria:2021hvt,Billo:2022fnb,Beccaria:2022ypy,Billo:2022lrv}. 
For later purposes, it is convenient to introduce the matrices $\mathsf{X}^{\mathrm{odd}}$ and
$\mathsf{X}^{\mathrm{even}}$ according to
\begin{equation}
    (\mathsf{X}^{\mathrm{odd}})_{k,\ell}\,\equiv\,\mathsf{X}_{2k+1,2\ell+1}  \quad\mbox{and}\quad
    (\mathsf{X}^{\mathrm{even}})_{k,\ell}\,\equiv\,\mathsf{X}_{2k,2\ell}
    \label{Xevenodd}
\end{equation}
and neglect the mixed even/odd elements $\mathsf{X}_{2k,2\ell+1}$ which will not play any role in the following. Using this notation, the interaction action (\ref{S0P}) becomes
\begin{align}
    S_0=-\frac{1}{2}\sum_{k,\ell=1}^\infty
    \cP_{2k+1}\,\mathsf{X}^{\mathrm{odd}}_{k,\ell}\,\cP_{2\ell+1}~.
    \label{S0P1}
\end{align}

As realized in \cite{Beccaria:2020hgy}, in the planar limit the vacuum expectation value of a product of many $\cP_k$'s in the Gaussian theory can be simply computed using Wick's theorem, with the contraction given in (\ref{PkPl0}). Thus, we can associate to each operator $\cP_{k}$ a real variable $p_k$ and write
\begin{align}
    \big\langle\cP_{k_1}\,\cP_{k_2}\ldots\cP_{k_n}\big\rangle_0=\int\! \cD\mathbf{p}
    ~p_{k_1}p_{k_2}\ldots p_{k_n}\,\rme^{-\frac{1}{2}\mathbf{p}^T\mathbf{p}}+O(N^{-1})
    \label{pk1pkn}
\end{align}
where $\mathbf{p}$ is a column vector with components $p_k$ and the integration measure is
\begin{align}
\cD\mathbf{p}=\prod_{k}\frac{dp_k}{\sqrt{2\pi}}~.
\end{align} 
Splitting $\mathbf{p}$ into its odd and even parts
$\mathbf{p}_{\mathrm{odd}}$ and $\mathbf{p}_{\mathrm{even}}$, whose components carry respectively odd and even indices, we have
\begin{align}
\mathbf{p}^T\,\mathbf{p}=\mathbf{p}^T_{\mathrm{even}}\,\mathbf{p}_{\mathrm{even}}+\mathbf{p}^T_{\mathrm{odd}}\,\mathbf{p}_{\mathrm{odd}}\qquad\mbox{and}\qquad
S_{0}=-\frac{1}{2}\,\mathbf{p}^T_{\mathrm{odd}}\,\mathsf{X}^{\mathrm{odd}}\,\mathbf{p}_{\mathrm{odd}}~,
\end{align}
while the integration measure clearly factorizes as $\cD\mathbf{p}=\cD\mathbf{p}_{\mathrm{even}}\,\cD\mathbf{p}_{\mathrm{odd}}$.
Therefore, in the planar limit the partition function of the \textbf{E}-theory becomes
\begin{align}
    \cZ_{\mathbf{E}}&=\int\!\cD\mathbf{p}_{\mathrm{even}}\,\rme^{-\frac{1}{2}\mathbf{p}^T_{\mathrm{even}}\,\mathbf{p}_{\mathrm{even}}}\int\,\cD\mathbf{p}_{\mathrm{odd}}\,\rme^{-\frac{1}{2}\mathbf{p}^T_{\mathrm{odd}}(\bone-\mathsf{X}^{\mathrm{odd}})\mathbf{p}_{\mathrm{odd}}}
    \,=\,\det\big(\bone-\mathsf{X}^{\mathrm{odd}}\big)^{-\frac{1}{2}}~,
\label{Z0}
\end{align}
corresponding to the free energy
\begin{align}
    \cF_{\mathbf{E}}=\frac{1}{2}\Tr\log\big(\bone-\mathsf{X}^{\mathrm{odd}}\big)~.
    \label{F0}
\end{align}
As mentioned above, this free energy has been studied in detail in \cite{Beccaria:2022ypy,Beccaria:2023kbl} where in particular its strong coupling behavior has been obtained using the asymptotic properties of the $\mathsf{X}$ matrix and the corresponding Bessel kernel.

In the following, we will need also the 3-point functions of the $\cP$ operators. As one can see from (\ref{pk1pkn}) with $n=3$, they vanish at the leading order in the large-$N$ expansion and become non-trivial only at the sub-leading order $1/N$. Furthermore, for parity reasons, these 3-point functions are non-zero only when there are either zero or two operators with odd indices. In these cases, one finds \cite{Billo:2022xas,Billo:2022fnb}:
\begin{subequations}
    \begin{align}
    \langle\cP_{2k}\,\cP_{2\ell}\,\cP_{2n}\rangle_0&=
    \frac{1}{N}\,d_{2k,2\ell,2n}~,\label{PPPeven}\\
    \langle\cP_{2k}\,\cP_{2\ell+1}\,\cP_{2n+1}\rangle_0&=
    \frac{1}{N}\,d_{2k,2\ell+1,2n+1}~,\label{PPPodd}
\end{align}
\label{PPP}%
\end{subequations}
where
\begin{align}
    d_{k,\ell,n}=\sqrt{k\,\ell\,n}~.
    \label{dkln}
\end{align}

Also the operators $\cM$ and $\cM^\prime$ in (\ref{Mis}) and (\ref{Mprimeis}) can be conveniently written in the $\cP$-basis. In plugging (\ref{avsP}) into (\ref{Mis}), we have to pay attention to the fact that both even and odd traces appear and, since $\big\langle \tr a^{2k}\big\rangle_0$ is non-zero, the resulting expression will contain terms with zero, one or two $\cP$'s. Therefore, after some algebra we can write
(see Appendix \ref{appendixM} for details)
\begin{align}
    \mathcal{M}=\mathcal{M}^{(0)}+\mathcal{M}^{(1)}+\mathcal{M}^{(2)}~,
    \label{M}
\end{align}
with
\begin{subequations}
    \begin{align}
    \mathcal{M}^{(0)}&=N^2\,\mathsf{M}_{0,0}+\mathsf{M}_{1,1}-\frac{1}{6}\,\sum_{k=1}^\infty\sqrt{2k+1}\,\mathsf{M}_{1,2k+1}+O(N^{-2})~,\label{M0}\\
    \mathcal{M}^{(1)}&=2N\sum_{k=1}^\infty \mathsf{M}_{0,2k}\,\cP_{2k}+O(N^{-1})~,\label{M1}\\
    \mathcal{M}^{(2)}&=\sum_{k,\ell=1}^\infty \Big[\mathsf{M}_{2k,2\ell}\,\cP_{2k}\cP_{2\ell}-\mathsf{M}_{2k+1,2\ell+1}\,\cP_{2k+1}\cP_{2\ell+1}\Big]~,\label{M2}
\end{align}
\label{M012}%
\end{subequations}
where the $\lambda$-dependent coefficients are given by the following convolutions of Bessel functions
\begin{subequations}
\begin{align}
    \mathsf{M}_{0,0}&=\int_0^\infty\!\frac{dt}{t}\,\frac{(t/2)^2}{\sinh(t/2)^2} \bigg[1-\frac{16\pi^2}{t^2\lambda}\,J_1\Big(\frac{t\sqrt{\lambda}}{2\pi}\Big)^2\bigg]~,\label{M00}\\[2mm]
    \mathsf{M}_{0,n}&=(-1)^{\frac{n}{2}+1}\,\sqrt{n}\!\int_0^\infty\!\frac{dt}{t}\,\frac{(t/2)^2}{\sinh(t/2)^2}\, \Big(\frac{4\pi}{t\sqrt{\lambda}}\Big)\,J_1\Big(\frac{t\sqrt{\lambda}}{2\pi}\Big)\,J_n\Big(\frac{t\sqrt{\lambda}}{2\pi}\Big)~,\label{M0n}\\[2mm]
    \mathsf{M}_{n,m}&=(-1)^{\frac{n+m+2nm}{2}+1}\,\sqrt{nm}\!\int_0^\infty\!\frac{dt}{t}\,\frac{(t/2)^2}{\sinh(t/2)^2}
    \,J_n\Big(\frac{t\sqrt{\lambda}}{2\pi}\Big)\,J_m\Big(\frac{t\sqrt{\lambda}}{2\pi}\Big)\label{Mnm}
\end{align}
\label{Mmatrix}%
\end{subequations}
for $n,m\geq 1$. Notice the similarity between $\mathsf{M}_{n,m}$ in (\ref{Mnm}) and $\mathsf{X}_{n,m}$ in (\ref{Xij}): the only difference is in the factor that multiplies the Bessel functions inside the integrand. Again, while the initial expression of $\mathcal{M}$ given in (\ref{Mis}) is a perturbative expansion in powers of the 't Hooft coupling, the final expressions (\ref{M012}) and (\ref{Mmatrix}) are exact in $\lambda$ and thus can be used also at strong coupling. The coefficient $\mathsf{M}_{0,0}$ which contributes to the planar limit of $\mathcal{M}$ agrees with the results of \cite{Russo:2013kea}, while the other coefficients $\mathsf{M}_{0,n}$ and $\mathsf{M}_{n,m}$ appear also in the study of the holographic correlators in $\cN=4$ SYM presented in \cite{Chester:2019pvm}.
The reason for this it that the first mass-correction of the $\cN=2^*$ theory is described by the operator $\mathcal{M}_0$ in (\ref{M0foot}) which is very similar to that of the $\mathbf{E}^*$-theory,
namely
\begin{align}
\mathcal{M}_0=\mathcal{M}_0^{(0)}+\mathcal{M}_0^{(1)}+\mathcal{M}_0^{(2)}
  \label{Mhat}
\end{align}
where
\begin{align}
    \mathcal{M}_0^{(0)}&=\mathcal{M}^{(0)}~,~\quad
    \mathcal{M}_0^{(1)}=\mathcal{M}^{(1)}~,~\quad
    \mathcal{M}_0^{(2)}=\sum_{k,\ell=1}^\infty \mathsf{M}_{k,\ell}\,\cP_{k}\,\cP_{\ell}~.
    \label{Mhat012}
\end{align}
The difference between $\mathcal{M}$ and $\mathcal{M}_0$ is only in the quadratic part and simply amounts to a different sign between the even and odd terms which however does not change the structure of the coefficients.

Let us now consider the operator $\cM^\prime$ in (\ref{Mprimeis}). Since this is a superposition of single traces, when we express it in the $\cP$-basis we only get terms with one $\cP$ or with no $\cP$. Thus, we can write
\begin{align}
    \cM^\prime=\cM^{\prime\,(0)}+\cM^{\prime\,(1)}~.
    \label{Mprime01}
\end{align}
By inserting (\ref{avsP}) into (\ref{Mprimeis}), one can immediately realize that
\begin{align}
    \cM^{\prime\,(1)}=O(N^0)\qquad\mbox{and}
\qquad    \cM^{\prime\,(0)}=O(N)~.
    \label{Mprime01a}
\end{align}
Indeed, in $\cM^{\prime\,(1)}$ the $N$-dependence disappears, while in $\cM^{\prime\,(0)}$ a net overall factor of $N$ remains in the product between $\langle \tr a^{2n}\rangle_0 $ (whose expression is given in (\ref{t2l})) and the terms accompanying the 't Hooft coupling in (\ref{Mprimeis}). 
By comparing $\cM^{\prime\,(0)}$ and $\cM^{\prime\,(1)}$ with $\cM^{(0)}$ and $\cM ^{ (1)}$ in (\ref{M0}) and (\ref{M1}), we see that the former have a power of $N$ lower than the latter. Therefore $\cM^\prime$ provides contributions that are sub-leading in the large-$N$ limit compared to those arising from $\cM$. For this reason from now on we will focus on $\cM$. This choice is clearly equivalent to assuming that the masses of the symmetric and anti-symmetric hypermultiplets are equal, {\it{i.e.}} $m_S=m_A=m$, which implies $m^\prime=0$.

\section{The first mass-correction to the free energy}
\label{sec:freeenergy}
In this section we describe the large-$N$ expansion of $\langle\,\mathcal{M}\,\rangle$, which according to (\ref{Fm1}) gives the first mass-correction to the free energy proportional to $m^2$. This calculation will also provide useful results for the analysis of the integrated correlators presented in the following sections.

From the explicit expression of the various components of $\mathcal{M}$ in (\ref{M012}), it is clear that we need to compute the 1- and 2-point functions of the operators $\mathcal{P}_k$ in the \textbf{E}-theory. These operators are normal-ordered and orthonormal in the Gaussian model but they are no longer so in the interacting matrix model of the \textbf{E}-theory, and thus their 1- and 2-point functions are non trivial.
Indeed, as shown in \cite{Billo:2022fnb,Billo:2022lrv}, one finds
\begin{subequations}
    \begin{align}
    \langle \cP_{2k}\,\cP_{2\ell}\rangle&=\delta_{k\ell}+O(N^{-2})~,\label{PPeveninE}\\[1mm]
    \langle \cP_{2k+1}\,\cP_{2\ell+1}\rangle&=(\mathsf{D}^{\mathrm{odd}})_{k,\ell}+O(N^{-2})
    \label{PPoddinE}
\end{align}
\label{PPinE}%
\end{subequations}
where\,%
\footnote{With an abuse of notation, we denote the inverse of a matrix $\mathsf{A}$ by $\frac{1}{\mathsf{A}}$.}
\begin{align}
    \mathsf{D}^{\mathrm{odd}}
    =\bone+\mathsf{X}^{\mathrm{odd}}+(\mathsf{X}^{\mathrm{odd}})^2+(\mathsf{X}^{\mathrm{odd}})^3+\ldots
    =\frac{1}{\bone-\mathsf{X}^{\mathrm{odd}}\phantom{\big|}}~.
    \label{Dnm}
\end{align}
Similarly one has
\begin{align}
    \langle \cP_{n}\rangle=0+O(N^{-1})
    \label{PinE}
\end{align}
where the appearance of sub-leading corrections is due to the fact that in the definition (\ref{avsP}) we used the vacuum expectation values of the Gaussian model and not those of the interacting theory.
As we shall see in the following, the $O(N^{-2})$-corrections in the 2-point functions (\ref{PPinE}) will not play any role since they will affect only sub-leading non-planar terms in the free energy that we will not consider. On the contrary, the $O(N^{-1})$-terms in the 1-point functions (\ref{PinE}) are important and cannot be neglected: they enter in the vacuum expectation value of $\mathcal{M}^{(1)}$, which has an explicit factor of $N$ in front, and thus contribute to the $O(N^0)$-part of the free energy.

Let us now proceed in order and compute the various contributions to $\langle\,\mathcal{M}\,\rangle$ in the large-$N$ expansion.
\paragraph{$\bullet$ $N^2$-terms:}
The only contribution proportional to $N^2$ comes from the first term in $\mathcal{M}^{(0)}$, and we simply have
\begin{align}
    \langle\, \mathcal{M}\, \rangle\big|_{N^2}=N^2\,\mathsf{M}_{0,0}
    \label{MN2}
\end{align}
with $\mathsf{M}_{0,0}$ given in (\ref{M00}). Note that this is the same result for the first mass-correction of the free energy in the $\mathcal{N}=2^*$ theory \cite{Russo:2013kea}. Therefore, to see some deviations in the $\mathbf{E}^*$-theory we have to consider the sub-leading terms.

\paragraph{$\bullet$ $N^1$-terms:} Contributions of order $N$ could originate from $\mathcal{M}^{(1)}$, but they contain $\langle\,\mathcal{P}_{2k}\,\rangle$. As one can see from (\ref{PinE}), this 1-point function is $O(N^{-1})$, and thus 
\begin{align}
    \langle\,\mathcal{M}\,\rangle\big|_{N}=0~.
    \label{MN1}
\end{align}
\paragraph{$\bullet$ $N^0$-terms:} The terms of order $N^0$ are interesting since they show a deviation from the $\mathcal{N}=2^*$ theory. They have three sources:
\begin{enumerate}
    \item The terms of order $N^0$ in $\langle\, \mathcal{M}^{(0)}\rangle$, namely
    \begin{align}
       \langle\,\mathcal{M}^{(0)}\rangle\big|_{N^0}= \mathsf{M}_{1,1}-\frac{1}{6}\sum_{k=1}^\infty \sqrt{2k+1}\,\mathsf{M}_{1,2k+1}~.
        \label{MN01}
    \end{align}
    \item The terms of order $N^0$ in $\langle\, \mathcal{M}^{(1)}\rangle$ which originate from the sub-leading correction in the 1-point function $\langle\mathcal{P}_{2k}\rangle$ (see (\ref{PinE})). As shown in Appendix \ref{appendixPka}, this 1-point function is
    \begin{align}
        \langle \mathcal{P}_{2k}\rangle=-\frac{\sqrt{2k}}{N}\,\lambda\,\partial_\lambda\cF_{\mathbf{E}}
        +O(N^{-3})
        \label{vevP2k}
    \end{align}
    where $\cF_{\mathbf{E}}$ is the free energy of the \textbf{E}-theory given in (\ref{F0}). Using this result, we find
    \begin{align}
        \langle\, \mathcal{M}^{(1)}\rangle\big|_{N^0}&=-2\lambda\,\partial_\lambda\cF_{\mathbf{E}}\,\sum_{k=1}^\infty \sqrt{2k}\,\mathsf{M}_{0,2k}=
        2\lambda\,\partial_\lambda\cF_{\mathbf{E}}\,\,\mathsf{M}_{1,1}
    \end{align}
    where in the last step we have exploited the relation (\ref{M02kM11}) that follows from the recurrence relations of the Bessel functions.
     \item The terms of order $N^0$ in $\langle\, \mathcal{M}^{(2)}\rangle$, which are
    \begin{align}
       \langle \,\mathcal{M}^{(2)}\rangle\big|_{N^0}= &\sum_{k,\ell=1}^\infty \Big[\mathsf{M}_{2k,2\ell}\,
        \langle \cP_{2k}\,\cP_{2\ell}\rangle-\mathsf{M}_{2k+1,2\ell+1}\,\langle \cP_{2k+1}\,\cP_{2\ell+1}\rangle\Big]\\
        &=\sum_{k=1}^\infty \mathsf{M}_{2k,2k}-\sum_{k,\ell=1}^\infty
        \mathsf{M}_{2k+1,2\ell+1}\,(\mathsf{D}^{\mathrm{odd}})_{k,\ell} 
        = \Tr \mathsf{M}^{\mathrm{even}}-\Tr \big(\mathsf{M}^{\mathrm{odd}}\,\mathsf{D}^{\mathrm{odd}}\big)\notag
    \end{align}
    where in the second line we have used (\ref{PPinE}) and introduced the even and odd parts of the matrix $\mathsf{M}$ in analogy with (\ref{Xevenodd}).
\end{enumerate}
Collecting all contributions, we obtain
\begin{align}
    \langle \,\mathcal{M}\,\rangle&= N^2\,\mathsf{M}_{0,0}\notag\\&~+\bigg[\big(1+2\lambda\,\partial_\lambda\cF_{\mathbf{E}}\big)\mathsf{M}_{1,1}-\frac{1}{6}\sum_{k=1}^\infty \sqrt{2k+1}\,\mathsf{M}_{1,2k+1}+\Tr \mathsf{M}^{\mathrm{even}}-\Tr \big(\mathsf{M}^{\mathrm{odd}}\,\mathsf{D}^{\mathrm{odd}}\big)\bigg]\label{vevMfin}\\
    &~+O(N^{-2})~.\notag
\end{align}
This formal expression is exact in $\lambda$. Indeed, all its terms are written using the matrix elements of $\mathsf{X}$ and $\mathsf{M}$, which are valid for all values of the 't Hooft coupling. 

For later convenience, we report also the result for first-mass correction of the free energy in the $\cN=2^*$ theory, namely $\langle \,\mathcal{M}_0\,\rangle_0$. From the expression of $\mathcal{M}_0$ given in (\ref{Mhat}), one easily finds
\begin{align}
    \langle \,\mathcal{M}_0\,\rangle_{0} &= N^2\,\mathsf{M}_{0,0}+\bigg[\mathsf{M}_{1,1}-\frac{1}{6}\sum_{k=1}^\infty \sqrt{2k+1}\,\mathsf{M}_{1,2k+1}+\Tr \mathsf{M}^{\mathrm{even}}+\Tr \mathsf{M}^{\mathrm{odd}}\bigg]
    +O(N^{-2})\notag\\
    &=N^2\,\mathsf{M}_{0,0}+\bigg[\sum_{k=1}^\infty
    \mathsf{M}_{k,k}-\frac{1}{6}\sum_{k=1}^\infty \sqrt{2k+1}\,\mathsf{M}_{1,2k+1}\bigg]
    +O(N^{-2})~.\label{vevMhatfin}
\end{align}
Notice that $\langle\,\mathcal{M}_0\,\rangle_0$ is not obtained from $\langle \,\mathcal{M}\,\rangle$ by simply turning-off the terms depending on the interaction action. Indeed, this would only reduce the interacting vacuum expectation values $\langle~\rangle$ to the free one $\langle~\rangle_0$ without changing $\cM$ into $\cM_0$.

\subsection{Weak coupling}
If $\lambda\to 0$ we can expand the Bessel functions inside the convolution integrals and easily compute the matrix elements $\mathsf{X}_{n,m}$ and $\mathsf{M}_{n,m}$ as power series in $\lambda$ whose coefficients are proportional to odd Riemann $\zeta$-values. Proceeding in this way, it is not difficult to show that
\begin{align}
    \mathsf{X}_{n,m}=O(\lambda^{\frac{n+m}{2}})~~~\mbox{and}~~~
    \mathsf{M}_{n,m}=O(\lambda^{\frac{n+m}{2}})~.
\end{align}
Therefore, if we are interested in obtaining the weak-coupling expansion up to a given perturbative order, we can systematically truncate the infinite matrices $\mathsf{X}$ and $\mathsf{M}$ to finite matrices and the computation of the infinite sums and of the traces appearing in (\ref{vevMfin}) is reduced to manipulations of finite expressions that can be effectively automated. Applying this method to (\ref{vevMfin}) at the first few perturbative orders, we have found
\begin{align}
  \langle \,\mathcal{M}\,\rangle &= N^2\,\Big[\,\frac{3\,
    \zeta_3}{2}\,\hat{\lambda}
    -\frac{25\,\zeta_5}{8}\,\hat{\lambda}^2
    +\frac{245\,\zeta_7}{32} \,\hat{\lambda}^3
    -\frac{1323\,\zeta_9}{64}\, \hat{\lambda}^4 
    + \frac{7623\,\zeta_{11}}{128}\, \hat{\lambda}^5
    +O(\hat{\lambda}^6)\,\Big] \notag\\[2mm]
    &~-\Big[\,\frac{3\,
    \zeta_3}{2}\,\hat{\lambda}
    -\frac{25\,\zeta_5}{8}\,\hat{\lambda}^2
    -\frac{175\,\zeta_7}{32} \,\hat{\lambda}^3
    +\frac{6615\,\zeta_9+360\,\zeta_3\,\zeta_5}{64}\, \hat{\lambda}^4 \notag\\[2mm]
    &\qquad- \frac{24255\,\zeta_{11}+1260\,\zeta_3\,\zeta_7+900\,\zeta_5^2}{32}\,\hat{\lambda}^5
    +O(\hat{\lambda}^6)\,\Big]+O(N^{-2})
    \label{weakE}
\end{align}
where $\hat{\lambda}=\lambda/(4\pi^2)$. These expansions
can actually be pushed to very high orders without much computational effort, and then used for numerical investigations.

In a similar way, we can obtain the weak-coupling expansion for the first mass-correction of the $\cN=2^*$ theory which is
\begin{align}
  \langle\, \mathcal{M}_0\,\rangle_0 &= N^2\,\Big[\,\frac{3\,
    \zeta_3}{2}\,\hat{\lambda}
    -\frac{25\,\zeta_5}{8}\,\hat{\lambda}^2
    +\frac{245\,\zeta_7}{32} \,\hat{\lambda}^3
    -\frac{1323\,\zeta_9}{64}\, \hat{\lambda}^4 
    + \frac{7623\,\zeta_{11}}{128}\, \hat{\lambda}^5
    +O(\hat{\lambda}^6)\,\Big] \notag\\[2mm]
    &~-\Big[\,\frac{3\,
    \zeta_3}{2}\,\hat{\lambda}
    -\frac{25\,\zeta_5}{8}\,\hat{\lambda}^2
    +\frac{245\,\zeta_7}{32} \,\hat{\lambda}^3
    -\frac{945\,\zeta_9}{64}\, \hat{\lambda}^4+O(\hat{\lambda}^6)\,\Big]+O(N^{-2})~.
    \label{weak2*}
\end{align}
Notice that in this expression, differently from (\ref{weakE}), all terms are linear in the Riemann $\zeta$-values and the difference with respect to the $\mathbf{E}^*$-theory starts in the $O(N^0)$-part at order $\hat{\lambda}^3$.

\subsection{Strong coupling}
The strong-coupling analysis of $\langle \,\mathcal{M}\,\rangle$ is less straightforward. The planar contribution, $N^2\,\mathsf{M}_{0,0}$, was studied in \cite{Russo:2013kea} where it was found that
\begin{align}
    \mathsf{M}_{0,0}~\underset{\lambda \rightarrow \infty}{\sim}
    \frac{\log\lambda}{2}+O(\lambda^0)~.
    \label{M00strong}
\end{align}
Let us now consider the $O(N^0)$-terms in the square brackets of (\ref{vevMfin}). The strong-coupling behavior of the matrix elements $\mathsf{M}_{n,m}$ can be derived by applying the Mellin-Barnes method.
Inserting the identity
\begin{align}
    J_n\Big(\frac{t\sqrt{\lambda}}{2\pi}\Big)\,J_m\Big(\frac{t\sqrt{\lambda}}{2\pi}\Big)=
    \int_{-\ii\infty}^{\ii\infty}\frac{ds}{2\pi\ii}\,\frac{\Gamma(-s)\,\Gamma(2s+n+m+1)}{\Gamma(s+n+1)\,\Gamma(s+m+1)\,\Gamma(s+n+m+1)}\,\Big(\frac{t\sqrt{\lambda}}{4\pi}\Big)^{2s+n+m}
\end{align}
in (\ref{Mnm}) and performing the integral over $t$, we get
\begin{align}
    \mathsf{M}_{n,m}&=(-1)^{\frac{n+m+2nm}{2}+1}\,\sqrt{nm}~\times\\
    &\quad\times \int_{-\ii\infty}^{\ii\infty}\frac{ds}{2\pi\ii}\,\frac{\Gamma(-s)\,\Gamma(2s+n+m+1)\,\Gamma(2s+n+m+2)\,\zeta_{2s+n+m+1}}{\Gamma(s+n+1)\,\Gamma(s+m+1)\,\Gamma(s+n+m+1)}\,\Big(\frac{\sqrt{\lambda}}{4\pi}\Big)^{2s+n+m}~.\notag
\end{align}
For large $\lambda$ we can close the integration contour over $s$ counter-clock wise and pick-up the residues over the poles on the negative real axis. In this way we find
\begin{align}
    \mathsf{M}_{n,m}~\underset{\lambda \rightarrow \infty}{\sim}
    -\frac{1}{2}\,\delta_{n,m}+\frac{\sqrt{n\,m}}{\sqrt{\lambda}}+O(\lambda^{-\frac{3}{2}})
    \label{M11strong}
\end{align}
for any $n,m>0$.
On the other hand, the analysis presented in \cite{Beccaria:2022ypy} shows that
\begin{align}
    \cF_{\mathbf{E}}~\underset{\lambda \rightarrow \infty}{\sim}\frac{\sqrt{\lambda}}{8}+O(\lambda^{0})~.
    \label{F0strong}
\end{align}
Therefore, we can easily conclude that the leading strong-coupling behavior of the first $O(N^0)$-term in (\ref{vevMfin}) is
\begin{align}
  \big(1+2\lambda\,\partial_\lambda\cF_{\mathbf{E}}\big)\mathsf{M}_{1,1}  ~\underset{\lambda \rightarrow \infty}{\sim}-\frac{\sqrt{\lambda}}{16}+O(\lambda^{0})~.
    \label{1sttermstrong}
\end{align}
Let us now consider the second $O(N^0)$-term in the square brackets of (\ref{vevMfin}). To find its strong-coupling behavior, we first make use of the identity (\ref{M12k+1}) and then use the Mellin-Barnes method to get
\begin{align}
   -\frac{1}{6}\,\sum_{k=1}^\infty\sqrt{2k+1}\,\mathsf{M}_{1,2k+1}&=-\frac{1}{12}\int_0^\infty \!\frac{dt}{t}\,\frac{(t/2)^2}{\sinh(t/2)^2} \,\Big(\frac{t \sqrt{\lambda }}{2 \pi }\Big)
    J_1\Big(\frac{t \sqrt{\lambda }}{2 \pi }\Big)\, J_2\Big(\frac{t \sqrt{\lambda }}{2   \pi }\Big)\notag\\
    &\underset{\lambda \rightarrow \infty}{\sim} -\frac{1}{24}+O(\lambda^{-\frac{1}{2}})~.
    \label{2ndtermstrong}
\end{align}
This term is sub-leading with respect to (\ref{1sttermstrong}) and thus does not contribute at leading order. 

We now analyze the last two terms in the square brackets of (\ref{vevMfin}).
Using the identities 
(\ref{TrModd}) and (\ref{TrMeven}), we can rewrite these terms as follows
\begin{align}
    \Tr \mathsf{M}^{\mathrm{even}}-\Tr \big(\mathsf{M}^{\mathrm{odd}}\,\mathsf{D}^{\mathrm{odd}}\big)&=
    \mathsf{M}_{1,1}+\frac{1}{2}\int_0^\infty\!\frac{dt}{t}\,\frac{(t/2)^2}{\sinh(t/2)^2}\,
    \Big(\frac{t \sqrt{\lambda }}{2 \pi }\Big)\,
     J_0\Big(\frac{t \sqrt{\lambda }}{2\pi }\Big)\, J_1\Big(\frac{t \sqrt{\lambda }}{2\pi }\Big)\notag\\[1mm]
     &~~-\Tr \Big[\mathsf{M}^{\mathrm{odd}}\,\big(\mathsf{D}^{\mathrm{odd}}-\bone\big)\Big]~.
     \label{3rdterm}
\end{align}
The strong-coupling behavior of the first line can be easily obtained using (\ref{M11strong}) and applying again the Mellin-Barnes method to the convolution of Bessel functions. In this way we find
\begin{align}
    \mathsf{M}_{1,1}+\frac{1}{2}\int_0^\infty\!\frac{dt}{t}\,\frac{(t/2)^2}{\sinh(t/2)^2}\,
    \Big(\frac{t \sqrt{\lambda }}{2 \pi }\Big)\,
     J_0\Big(\frac{t \sqrt{\lambda }}{2\pi }\Big)\, J_1\Big(\frac{t \sqrt{\lambda }}{2\pi }\Big)
\underset{\lambda \rightarrow \infty}{\sim} -\frac{1}{4} +O(\lambda^{-\frac{1}{2}})~.
\end{align}
Also this contribution is sub-leading with respect to (\ref{1sttermstrong}) and can be neglected at leading order. 

We thus remain with the term in the second line of (\ref{3rdterm}). Its strong-coupling behavior cannot be deduced from a straightforward application of the Mellin-Barnes method due to the presence of (infinitely) many terms with multiple insertion of $\mathsf{X}^{\mathrm{odd}}$ arising from the interaction action. However, the leading term at strong coupling can be obtained by applying the techniques developed in \cite{Belitsky:2020qrm,Belitsky:2020qir} for the study of the octagon form factor of $\cN=4$ SYM in the strong-coupling regime.
In Appendix~\ref{app:strong} we provide some details on these techniques and show that
\begin{align}
 -\Tr \Big[\mathsf{M}^{\mathrm{odd}}\,\big(\mathsf{D}^{\mathrm{odd}}-\bone\big)\Big]\underset{\lambda \rightarrow \infty}{\sim}
 -\frac{\sqrt{\lambda}}{24}+O(\lambda^0)~.
 \label{3rtermstrong}
\end{align}
Collecting all contributions, from (\ref{M00strong}), (\ref{1sttermstrong}) and
(\ref{3rtermstrong}) we conclude that
\begin{align}
     \langle \,\mathcal{M}\,\rangle~
     \underset{\lambda \rightarrow \infty}{\sim} N^2\,\frac{\log\lambda}{2}
 -\frac{5\sqrt{\lambda}}{48}+O(N^{-2})~.
 \label{strongE}
\end{align}
We observe that this strong-coupling behavior is similar but not identical to that of the free energy in the $\cN=2^*$ theory which is given by \cite{Russo:2013kea}
\begin{align}
     \langle \,\mathcal{M}_0\,\rangle_{0}~
     \underset{\lambda \rightarrow \infty}{\sim}
     N^2\,\frac{\log\lambda}{2}
 -\frac{\sqrt{\lambda}}{6}+O(N^{-2})~.
 \label{strong2*}
\end{align} 
The difference between (\ref{strongE}) and (\ref{strong2*}) in the $O(N^0)$-terms signals the non-planar inequivalence of the $\cN=2^*$ and $\mathbf{E}^*$ theories also at leading order in the strong-coupling expansion.

Before concluding this section, for completeness we briefly mention that if we had chosen different values for the hypermultiplet masses of the $\mathbf{E}^*$-theory, we would have found an $O(N)$-contribution to the free energy proportional to $m^{\prime\,2}$ coming from $\langle \,\mathcal{M}^{\prime\,(0)}\,\rangle$ (see (\ref{Mprime01a})). At large $N$, however, this is sub-leading with respect to the contribution proportional to $m^2$ which is $O(N^2)$ and comes from $\langle \,\mathcal{M}^{(0)}\,\rangle$. Thus, to obtain the first mass-correction of the free energy of the $\mathbf{E}^*$-theory at leading order in the large-$N$ expansion, one can neglect $\cM^\prime$ and just consider $\cM$.
On the other hand, as we will see, the manipulations we performed to obtain the sub-leading $O(N^0)$-terms of $\langle \,\mathcal{M}\,\rangle$, are extremely useful for the integrated correlators to which we now turn.

\section{Integrated correlators: a first example}
\label{sec:integrated}
In this section we begin a detailed study of the quantity introduced in (\ref{derivativeZE}), namely
\begin{align}
    \partial_{\tau_p} \partial_{\,\overbar{\tau}_p}\partial_m^2 \log\cZ_{\mathbf{E}^*}\Big|_{m=0}\,\equiv\,\cC_p~.
    \label{CpEis}
\end{align}
However, as a warm-up we first review what happens in $\cN=4$ SYM considering
\begin{align}
    \partial_{\tau_p} \partial_{\,\overbar{\tau}_p}\partial_m^2 \log\cZ_{\cN=2^*}\Big|_{m=0}\,\equiv\,\cC^{(0)}_p
    \label{Cp4is}
\end{align}
for $p=2$.
This exercise is useful because it allows us to easily retrieve the results of \cite{Binder:2019jwn} using our approach, paving the way for the generalization to the \textbf{E}-theory.

\subsection{\texorpdfstring{$p=2$}{} in \texorpdfstring{$\cN=4$}{} SYM }
When $p=2$ we can take advantage of the standard complexified gauge coupling
\begin{align}
    \tau=\frac{\theta}{2\pi}+\ii\,\frac{4\pi N}{\lambda}\label{tau}~,
\end{align}
where $\theta$ is the topological vacuum angle and show that \cite{Chester:2019jas}
\begin{align}
    \cC_2^{(0)}=\frac{1}{4}\,\Delta_\tau\partial_m^2\log\cZ_{\cN=2^*}\Big|_{m=0}~.
    \label{C20}
\end{align}
Here we have adopted the same conventions as in \cite{Dorigoni:2021bvj,Dorigoni:2021guq},
and introduced the hyperbolic Laplacian $\Delta_\tau=4 (\mathrm{Im}\,\tau)^2\,\partial_\tau\partial_{\overbar{\tau}}$. 
On the other hand, exploiting the localization results presented in Sections~\ref{sec:Estar} and \ref{sec:freeenergy} we can write
\begin{align}
\cC_2^{(0)}=\frac{1}{2}\,\Delta_\tau \langle\,\mathcal{M}_0\,\rangle_{0}
    \label{C20a}
\end{align}
where $\mathcal{M}_0$ is the operator defined in (\ref{M0foot}) and whose vacuum expectation value is given in (\ref{vevMhatfin}) in the large-$N$ expansion.
Since the topological vacuum angle is set to zero, we can express the hyperbolic Laplacian only in terms of derivatives with respect to the 't Hooft coupling according to $\Delta_\tau=(2\lambda\partial_\lambda+\lambda^2\partial_\lambda^2)$. Hence, we have
\begin{align}
    \cC_2^{(0)}=\frac{1}{2}\big(2\lambda\,\partial_\lambda+\lambda^2\,\partial_\lambda^2\big)\langle\,\mathcal{M}_0\,\rangle_{0}~.
    \label{C20b}
\end{align}
Using (\ref{vevMhatfin}) and the explicit form of the coefficients $\mathsf{M}_{n,m}$ in terms of Bessel functions, one can verify that this expression exactly reproduces the results of \cite{Dorigoni:2021bvj,Dorigoni:2021guq} in the large-$N$ expansion. 

We now observe that
\begin{align}
    \lambda\partial_\lambda \langle\,\mathcal{M}_0\,\rangle_{0}&=
    \langle\tr a^2 \mathcal{M}_0\,\rangle_{0}-
    \langle\tr a^2\rangle_{0}\,
    \langle\,\mathcal{M}_0\,\rangle_{0}~.
    \label{dlambda}
\end{align}
This relation can be proved by writing $\langle\,\mathcal{M}_0\,\rangle_{0}$ as a matrix integral over $a$ and performing the inverse rescaling (\ref{rescaling}) in such a way that all $\lambda$-dependence is removed from $\mathcal{M}_0$ and put entirely in the Gaussian term in the exponent and in a Jacobian prefactor produced by the integration measure. In this way one can realize that a $\lambda$-derivative can be traded for an insertion of $\tr a^2$ in the matrix integral coming from the Gaussian term, and an insertion of $\langle\tr a^2\rangle_{0}=(N^2-1)/2$ coming from the Jacobian. If we now introduce the operator
\begin{align}
    O_2^{(0)}=\tr a^2-\langle\tr a^2\rangle_{0}=\sqrt{\frac{N^2}{2}}\,\cP_2~,
    \label{O20}
\end{align}
we can rewrite (\ref{dlambda}) simply as
\begin{align}
    \lambda\partial_\lambda \langle\,\mathcal{M}_0\,\rangle_{0}=
    \langle O_2^{(0)} \mathcal{M}_0\,\rangle_{0}~.
    \label{dlambda1}
\end{align}
The operator $O_2^{(0)}$ defined in (\ref{O20}) is the matrix-model counterpart of the chiral and anti-chiral operators $\cO_2(x)$ and $\overbar{\cO}_2(x)$ of the $\cN=4$ SYM, and its 2-point correlator yields the normalization coefficient $\cG_2^{(0)}$ of the 2-point function introduced in (\ref{norm2point}), namely
\begin{align}
    \cG_2^{(0)}=\langle O_2^{(0)} O_2^{(0)}\rangle_{0}=\frac{N^2}{2}\,
    \langle\cP_2\, \cP_2\rangle_{0}=\frac{N^2}{2}
    \label{G20}
\end{align}
where in the last step we have used (\ref{PkPl0}). 

Proceeding along the same lines, one can show that
\begin{align}
    \lambda^2\partial^2_\lambda \langle\,\mathcal{M}_0\,\rangle_{0}=
    \langle O_2^{(0)} O_2^{(0)} \mathcal{M}_0\,\rangle_{0}
    -\langle O_2^{(0)}O_2^{(0)}\rangle_{0}\,\langle\,\mathcal{M}_0\,\rangle_{0}-2\,\langle O_2^{(0)}\mathcal{M}_0\,\rangle_{0}~.
    \label{dlambda2}
\end{align}
Therefore, inserting (\ref{dlambda1}) and (\ref{dlambda2}) into (\ref{C20b}) we find
\begin{align}
    \cC_2^{(0)}=\frac{1}{2}\,\langle O_2^{(0)} O_2^{(0)} \mathcal{M}_0\,\rangle_{0}
    -\frac{1}{2}\,\langle O_2^{(0)} O_2^{(0)}\rangle_{0}\,\langle\,\mathcal{M}_0\,\rangle_{0}
    \,\equiv\,\frac{1}{2}\,
    \vvev{O_2^{(0)} O_2^{(0)} \mathcal{M}_0\,}_0
   ~.
   \label{C20fin}
\end{align}
This result clarifies the meaning of the operations to be performed on the logarithm of the partition function to obtain $\cC_2^{(0)}$. Indeed, the $\tau$-derivatives produce two insertions of $O_2^{(0)}$ which is the matrix-model representative of the Coulomb-branch operators with dimension 2 of the gauge theory, while the mass-derivatives associated to the integrated insertions of two moment-map operators correspond effectively to the insertion of $\mathcal{M}_0$ inside a ``connected'' correlator that we have denoted with the symbol $\langle\hspace{-2pt}\langle ~\rangle\hspace{-2pt}\rangle_{0}$. One can therefore conclude that the integrated 4-point function of two Coulomb-branch operators with dimension 2 and two moment-map operators in $\cN=4$ SYM is captured in the matrix model by the 3-point connected correlator defined in the right-hand side of (\ref{C20fin}).

\subsection{\texorpdfstring{$p=2$}{} in \textbf{E}-theory} 
The previous analysis can be easily extended to the \textbf{E}-theory. In this case we have of course to replace $\mathcal{M}_0$ with the operator $\mathcal{M}$ defined in (\ref{Mis}) and also pay attention to the fact that the Coulomb-branch operators $\cO_2(x)$ and $\overbar{\cO}_2(x)$ of the \textbf{E}-theory are not any more represented in the matrix model by $O_2^{(0)}$. Indeed, the latter is normal-ordered with respect to the free Gaussian matrix model but not with respect to the interacting matrix model so that one has to subtract its expectation value in the \textbf{E}-theory vacuum. Therefore, the appropriate dimension-2 operator in the \textbf{E}-theory is
\begin{align}
    O_2=\sqrt{\cG_2^{(0)}}\,\big(\cP_{2}-\langle\cP_{2}\rangle\big)~.
    \label{O2is}
\end{align}
Using this definition we find
\begin{align}
  \cG_2=  \langle O_2\, O_2\rangle=\cG_2^{(0)}\big(\langle\cP_2\,\cP_2\rangle
  -\langle\cP_2\rangle^2\big) = \cG_2^{(0)}\big[1+O(N^{-2})\big]
  \label{G2is}
\end{align}
where in the last step we have used (\ref{PPeveninE}) and taken into account the the 1-point functions are $O(N^{-1})$ as shown in (\ref{vevP2k}). We thus retrieve a well-known fact, namely that $\cG_2$ and $\cG_2^{(0)}$ are the same in the planar limit \cite{Billo:2022xas}.

Following the same steps described above for $\cN=4$ SYM, one can show quite straightforwardly that
\begin{align}
    \cC_2=\frac{1}{2}\,\langle O_2\, O_2\, \mathcal{M}\,\rangle
    -\frac{1}{2}\,\langle O_2 \,O_2\rangle\,\langle\,\mathcal{M}\,\rangle
    \,\equiv\,\frac{1}{2}\,\vvev{O_2\,O_2\,\mathcal{M}\,}~.
   \label{C2fin}
\end{align}
This result provides the matrix-model representation of the mixed integrated correlator among two Coulomb-branch operators of dimension 2 and two moment-map operators in the \textbf{E}-theory.

\section{Integrated correlators: the general case}
\label{sec:integratedp}
We now study the integrated correlators for generic $p$, starting again from $\cN=4$ SYM.

\subsection{Integrated correlators in \texorpdfstring{$\cN=4$}{} SYM}
Let us consider the quantity $\cC_p^{(0)}$ defined in (\ref{Cp4is}). As discussed in \cite{Gerchkovitz:2016gxx}, in the matrix model the derivatives with respect to the couplings $\tau_p$, and $\bar\tau_p$ produce insertions of normal-ordered operators $O_p^{(0)}$ which correspond to the Coulomb-branch operators $\cO_p(x)$ and $\overbar{\cO}_p(x)$ of the gauge theory. Thus, in analogy with the $p=2$ case in (\ref{C20fin}), we have
\begin{align}
    \cC_p^{(0)}=\frac{1}{2}\,\vvev{O_p^{(0)}\,O_p^{(0)}\,\mathcal{M}_0\,}_{0}~.
\label{Cp4mm}
\end{align}
In the large-$N$ limit the operator $O_p^{(0)}$ is given by 
\begin{align}
    \label{OtcP}
        O_p^{(0)} = \sqrt{\cG_p^{(0)}}\, \cP_p~,
\end{align}
with $\cG_p^{(0)} = p\,(N/2)^p$.
Indeed, at large $N$ the $\cP$-operators have canonical a 2-point functions up to sub-leading corrections, see (\ref{PkPl0}), and the $\cG_p^{(0)}$ factor restores the standard normalization of the Coulomb-branch operators of the field theory. 

In large-$N$ limit we can use in (\ref{Cp4mm}) the explicit expression of $\mathcal{M}_0$ given in (\ref{Mhat}) and (\ref{Mhat012}), and obtain
up to sub-leading terms 
\begin{align}
        \frac{\cC_p^{(0)}}{\cG_p^{(0)}} & =
        N\sum_{k=1}^\infty \mathsf{M}_{0,2k}\,\langle \cP_p\,\cP_p\, \cP_{2k}\rangle_0 
        +\frac{1}{2} \sum_{k,\ell=1}^\infty \mathsf{M}_{k,\ell}\,\big(\langle\cP_p\,\cP_p\,\cP_k\,\cP_\ell\rangle_0- \langle \cP_p\,\cP_p\rangle_0\,\langle\cP_k\,\cP_\ell\rangle_0\big)~.
        \label{ppM01}
\end{align}
In the first sum we can insert the value of the 3-point correlators (\ref{PPP}), while in the second sum we can evaluate the 4-point correlator using Wick's theorem which reduces it to product of 2-point correlators. In this way we find
\begin{align}
        \frac{\cC_p^{(0)}}{\cG_p^{(0)}}=\,
        p\sum_{k=1}^\infty \sqrt{2k}\,\mathsf{M}_{0,2k} +\mathsf{M}_{p,p}\,=\,
        \mathsf{M}_{p,p}-p\,\mathsf{M}_{1,1}
        \label{ppM02}
\end{align}
where the last step follows from the identity (\ref{M02kM11}). 
If we use the definition (\ref{Mnm}) of the matrix elements $\mathsf{M}_{n,m}$ we easily realize that $\cC_p^{(0)}$ is a convolution integral of Bessel functions proportional to $(J_p^2-J_1^2)$
in perfect agreement with the results of \cite{Binder:2019jwn}.

From (\ref{ppM02}) we can also extract the leading strong-coupling behavior of the integrated correlator in a straightforward way. Indeed, exploiting (\ref{M11strong}) we have
\begin{align}
    \frac{\cC_p^{(0)}}{\cG_p^{(0)}} ~\underset{\lambda \rightarrow \infty}{\sim}\,\frac{p-1}{2}+O(\lambda^{-\frac{1}{2}})
    \label{Cp0strong}
\end{align}
which is precisely equation (\ref{ratioN4}).

\subsection{Integrated correlators in \textbf{E}-theory}
In the \textbf{E}-theory we have to consider the correlators
\begin{align}
    \label{intcE}
        \cC_p = \frac 12\, \vvev{O_p\, O_p \,\mathcal{M}\,}
\end{align}
and distinguish the cases in which $p$ is even or odd.

\subsubsection*{Even operators}
When $p = 2q$, the calculation is similar to that of $\cN=4$ SYM. Indeed, the operators $O_{2q}$ are as in (\ref{OtcP})
except for the subtraction of their expectation value, namely
\begin{align}
    O_{2q}=\sqrt{\cG_{2q}^{(0)}}\,\big(\cP_{2q}-\langle\cP_{2q}\rangle\big)~.
    \label{O2qis}
\end{align}
From this we immediately find
\begin{align}
  \cG_{2q}=  \langle O_{2q}\, O_{2q}\rangle=\cG_{2q}^{(0)}\big(\langle\cP_{2q}\,\cP_{2q}\rangle-\langle\cP_{2q}\rangle^2\big) = \cG_{2q}^{(0)}\,\big[1+O(N^{-2})\big]~.
  \label{G2qis}
\end{align}
These two equations are the obvious generalizations of (\ref{O2is}) and (\ref{G2is}).

Inserting (\ref{O2qis}) into (\ref{intcE}) and using the large-$N$ expansion of the operator $\cM$ given in 
(\ref{M}) and (\ref{M012}), we have
\begin{align}
    \label{C2qis}
        \frac{\cC_{2q}}{\cG_{2q}} & = N \sum_{k=1}^\infty \mathsf{M}_{0,2k}\, \vvev{O_{2q}\, O_{2q}\, \cP_{2k}}+ \frac 12 \sum_{k,\ell=1}^\infty \mathsf{M}_{2k,2\ell} \,\vvev{O_{2q}\, O_{2q}\, \cP_{2k} \,\cP_{2\ell}}
        \notag\\
        &\quad
        - \frac 12 \sum_{k,\ell=1}^\infty \mathsf{M}_{2k+1,2\ell+1} \vvev{O_{2q}\, O_{2q}\, \cP_{2k+1} \cP_{2\ell+1}}~.
\end{align}
The fact that the $\langle\hspace{-2pt}\langle~\rangle\hspace{-2pt}\rangle$ correlators are $\cM$-connected implies that there must be some non-trivial contraction between the $O$ and the $\cP$ operators. This has the immediate consequence that the last term in (\ref{C2qis}) vanishes in the planar limit due to the different parities of the $O$ and $\cP$ operators. We thus have only to consider the first line of (\ref{C2qis}).
The correlator in the first term is
\begin{align}
    \vvev{O_{2q}\, O_{2q}\, \cP_{2k}}=
    \vvev{\big(\cP_{2q}- \vev{\cP_{2q}}\big)\big(\cP_{2q}- \vev{\cP_{2q}}\big)
        \cP_{2k}} =\langle \cP_{2q}\,\cP_{2q}\,\cP_{2k}\rangle_{\mathrm{conn}}
\end{align}
Indeed, by removing the contributions that are disconnected from the operator $\cP_{2k}$ coming from $\mathcal{M}$, we simply obtain the connected expectation value of three $\cP$ operators. 
When all these operators are even, the interaction action of the $\mathbf{E}$-theory matrix model does not play any role and the analogue of (\ref{PPPeven}) holds, namely
\begin{align}
    \label{deee}
        \langle \cP_{2q}\,\cP_{2r}\,\cP_{2s}\rangle_{\mathrm{conn}}= \frac 1N d_{2q,2r,2s} = \frac{1}{N}\,\sqrt{(2q)(2r)(2s)}~.
\end{align}
Thus, we have
\begin{align}
    \vvev{O_{2q}\, O_{2q}\, \cP_{2k}}=\frac{1}{N}\,2q\,\sqrt{2k}~.
    \label{O2q2q2k}
\end{align}
The remaining correlator to be considered in (\ref{C2qis}) is
\begin{align}
    \label{C2is1}
    \vvev{O_{2q}\, O_{2q}\, \cP_{2k} \,\cP_{2\ell}}=\vvev{ \big(\cP_{2q}- \vev{\cP_{2q}}\big)\big(\cP_{2q}- \vev{\cP_{2q}}\big)
        \cP_{2k}\,\cP_{2\ell}}~.
\end{align}
Here the leading terms in the Wick decomposition at large $N$ are given by the product of two 2-point functions of $\cP$ operators, which are of order $N^0$; instead, the terms involving the 1-point functions are sub-leading since $\langle\cP_{2q}\rangle$ is $O(N^{-1})$ as we have seen in (\ref{vevP2k}). Thus, in the planar limit (\ref{C2is1}) reduces thus to
\begin{align}
    \label{C2is3}
    \vvev{O_{2q}\, O_{2q}\, \cP_{2k} \,\cP_{2\ell}}=
        2\,\langle\cP_{2q}\,\cP_{2k}\rangle \, \langle \cP_{2q}\,\cP_{2\ell}\rangle  = 2\,\delta_{q,k} \,\delta_{q,\ell} ~.
\end{align}

Inserting these findings into (\ref{C2qis}) we get
\begin{align}
    \label{C2is4}
        \frac{\cC_{2q}}{\cG_{2q}} = 2q \sum_{k=1}^\infty \mathsf{M}_{0,2k} \sqrt{2k} + \mathsf{M}_{2q,2q} = \mathsf{M}_{2q,2q} - 2 q \,\mathsf{M}_{1,1}~.
\end{align}
This result is identical to that of $\cN=4$ SYM given in (\ref{ppM02}). At strong coupling, using (\ref{M11strong}), we easily see that
\begin{align}
    \frac{\cC_{2q}}{\cG_{2q}} ~\underset{\lambda \rightarrow \infty}{\sim}\,\frac{2q-1}{2}+O(\lambda^{-\frac{1}{2}})
    \label{C2qstrong}
\end{align}
which is precisely equation (\ref{ratioE}) for $p=2q$.

\subsubsection*{Odd operators}
When $p=2q+1$ there are some important differences. First of all,
the correlators of two odd $\cP$'s are not diagonal in the \textbf{E}-theory (see (\ref{PPoddinE})). This fact implies that the matrix-model operators $O_{2q+1}$ representing the Coulomb-branch operators $\cO_{2q+1}(x)$ and $\overbar{\cO}_{2q+1}(x)$ of the gauge theory, which have diagonal 2-point functions, must be obtained from the odd $\cP$'s by a Gram-Schmidt procedure with the result that 
\begin{align}
    \label{cPodd1}
        O_{2q+1} = \sqrt{\cG_{2q+1}^{(0)}} \Big(\cP_{2q+1} - \sum_{q^\prime<q} \mathsf{Q}_{q,q^\prime} \cP_{2q^\prime+1}\Big)~.
\end{align}
The coefficients $\mathsf{Q}_{q,q^\prime}$ are determined by demanding that $O_{2q+1}$ be orthogonal to all operators of lower dimension and, as shown in \cite{Beccaria:2020hgy,Beccaria:2021hvt,Billo:2022xas}, can be explicitly expressed in terms of the coefficients $(\mathsf{D}^{\mathrm{odd}})_{q,r}$ appearing in the 2-point functions (\ref{Dnm}).

Using the expansion (\ref{cPodd1}), we see that the odd integrated correlators
\begin{align}
    \label{Co1}
        \cC_{2q+1}= \frac 12 \vvev{O_{2q+1} O_{2q+1} \cM}
\end{align}
are determined by the quantities
\begin{align}
    \label{Co2}
        \Pi_{q,r} = \vvev{\cP_{2q+1}\,\cP_{2r+1}\,\cM}~.
\end{align}
In fact, one has
\begin{align}
    \label{Co2bis}
        \frac{\cC_{2q+1}}{\cG_{2q+1}^{(0)}} = \frac{1}{2}\,\Big(\Pi_{q,q} - 2\sum_{q^\prime<q} \mathsf{Q}_{q,q^\prime} \Pi_{q^\prime,q} 
        + \sum_{q^\prime,q^{\prime\prime}<q} \mathsf{Q}_{q,q^\prime} \Pi_{q^\prime,q^{\prime\prime}}\mathsf{Q}_{q,q^{\prime\prime}}\Big)~. 
\end{align}
Let us then consider in more detail the coefficients $\Pi_{q,r}$.
Using arguments similar to those employed in the even case, we see that the fact that the ``external'' operators $\cP_{2q+1}$ and $\cP_{2r+1}$ in (\ref{Co2}) have to be connected with $\cM$ implies that 
this time it is only the odd part of $\cM^{(2)}$ that matters. Indeed, we find
\begin{align}
    \label{Co4}
        \Pi_{q,r} = 2 N \sum_{k=1}^\infty \mathsf{M}_{0,2k} \vvev{\cP_{2q+1}\,\cP_{2r+1}\,\cP_{2k}} - \sum_{k,\ell=1}^\infty \mathsf{M}_{2k+1,2\ell+1} \,\vvev{\cP_{2q+1}\,\cP_{2r+1}\,\cP_{2k+1}\,\cP_{2l+1}}~.
\end{align}
Since the $\vvev$ correlators are $\cM$-connected, in order to have a non-vanishing contribution there must be a contraction between the external operators and those whose indices are summed over. Thus in the large-$N$ limit upon using Wick's theorem we get\,%
\footnote{Notice that the term $\vev{\cP_{2q+1}\,\cP_{2r+1}} \sum_{k,\ell=1}^\infty \mathsf{M}_{2k+1,2\ell+1} \,\vev{\cP_{2k+1}\,\cP_{2l+1}}$ cancels in a $\cM$-connected correlator.}
\begin{align}
    \label{Co5}
        \Pi_{q,r} & = 2N \sum_{k=1}^\infty \mathsf{M}_{0,2k} \vev{\cP_{2q+1}\,\cP_{2r+1}\,\cP_{2k}}_{\mathrm{conn}} 
        - 2\!\sum_{k,\ell=1}^\infty \mathsf{M}_{2k+1,2\ell+1} \,\vev{\cP_{2q+1}\,\cP_{2k+1}} \vev{\cP_{2r+1}\,\cP_{2l+1}}~.
\end{align}
The connected 3-point correlators appearing in the first sum above have been studied in detail in \cite{Billo:2022fnb} where it was shown that
\begin{align}
    \label{Co3}
        \vev{\cP_{2q+1}\,\cP_{2r+1}\,\cP_{2k}}_{\mathrm{conn}} = \frac 1N \,\mathsf{d}_{2q+1}\,\mathsf{d}_{2r+1}\sqrt{2k} + O(N^{-2})
\end{align}
with
\begin{align}
    \label{dis}
        \mathsf{d}_{2q+1} =
        \sum_{q^\prime=1}^\infty (\mathsf{D}^{\rm odd})_{q,q^\prime}\,\sqrt{2q^\prime + 1}~.
\end{align}
Using this result and the 2-point correlators (\ref{PPoddinE}), we can rewrite (\ref{Co5}) as 
\begin{align}
    \Pi_{q,r}&=2 \,\mathsf{d}_{2q+1},\mathsf{d}_{2r+1} \sum_{k=1}^\infty \mathsf{M}_{0,2k} \sqrt{2k} 
        - 2 \sum_{k,\ell=1}^\infty 
        \mathsf{M}_{2k+1,2\ell+1} \,(\mathsf{D}^{\mathrm{odd}})_{q,k}\,(\mathsf{D}^{\mathrm{odd}})_{\ell,r}\notag\\[1mm]
        &=-2\,\mathsf{d}_{2q+1}\,\mathsf{d}_{2r+1}\,\mathsf{M}_{1,1} -2\,\big(\mathsf{D}^{\mathrm{odd}} \,\mathsf{M}^{\mathrm{odd}} \,\mathsf{D}^{\mathrm{odd}}\big)_{q,r}
        \label{Co6}
\end{align}
where in the second line we have used once again the identity (\ref{M02kM11}).
Inserting this result into (\ref{Co2bis}), the expression of $\cC_{2q+1}$ can be worked out in principle for any value of $q$. It simplifies a lot for low values of $q$, and specifically for $q=1$ when it reduces to
\begin{align}
    \cC_3=\frac{1}{2}\,\cG_3^{(0)}\,\Pi_{1,1}
    ~.
    \label{C3}
\end{align}
It also simplifies for arbitrary $q$ in the strong-coupling limit $\lambda\to\infty$ in which we are particularly interested. Let us then give some details for this case.

At strong coupling the expression (\ref{cPodd1}) for $O_{2q+1}$ becomes \cite{Billo:2022fnb} 
\begin{align}
    \label{Co9}
        O_{2q+1} ~\underset{\lambda \rightarrow \infty}{\sim}\,\sqrt{\cG^{(0)}_{2q+1}} \Big(\cP_{2q+1} - \sqrt{\frac{2q+1}{2q-1}}\,\cP_{2q-1}\Big)~.
\end{align}
It then follows that
\begin{align}
    \label{Co10}
        \frac{\cC_{2q+1}}{\cG^{(0)}_{2q+1}}~\underset{\lambda \rightarrow \infty}{\sim}~
        \frac 12 \,\Big(\Pi_{q,q} - 2 \,\sqrt{\frac{2q+1}{2q-1}}\,\Pi_{q,q-1} + \frac{2q+1}{2q-1}\, \Pi_{q-1,q-1}\Big)
\end{align}
where in the right-hand side we obviously must use the asymptotic form of the $\Pi$-coefficients for $\lambda\to\infty$. To find this, let us consider the two terms in (\ref{Co6}). The first term is under full control analytically, since in \cite{Billo:2022fnb} it was shown that
\begin{align}
    \label{Co11}
        \mathsf{d}_{2q+1} ~\underset{\lambda \rightarrow \infty}{\sim}~\frac{2\pi}{\sqrt{\lambda}}\, \sqrt{2q+1}\,q\,(q+1)
\end{align}
and $\mathsf{M}_{1,1}$ behaves as stated in (\ref{M11strong}). Thus, altogether we have
\begin{align}
    \label{Co12}
        -2 \,\mathsf{d}_{2q+1}\,\mathsf{d}_{2r+1}\, \mathsf{M}_{1,1} 
        ~\underset{\lambda \rightarrow \infty}{\sim}~
        \frac{4\pi^2}{\lambda} \sqrt{(2q+1)}\, q\,(q+1)\,\sqrt{2r+1}\, r\,(r+1) + O\big(\lambda^{-\frac{3}{2}}\big)~.
\end{align}
As for the second term in (\ref{Co6}), we did not succeed in deriving its large-$\lambda$ behavior in an analytic way. However, we have studied it numerically for the lowest values of $q$ and $r$ and found that
\begin{align}
\label{Co13}
    \big(\mathsf{D}^{\mathrm{odd}} \mathsf{M}^{\mathrm{odd}} \mathsf{D}^{\mathrm{odd}}\big)_{q,r}
     ~\underset{\lambda \rightarrow \infty}{\sim}~ O(\lambda^{-\frac{3}{2}})
\end{align}
(see Appendix~\ref{app:strongDMD} for details).
Based on this numerical evidence, we conjecture that the behavior (\ref{Co13}) holds for generic $q$ and $r$ so that the second term in (\ref{Co6}) can be neglected with respect to the first one being sub-leading at strong coupling. Thus, we can use the approximation
\begin{align}
    \Pi_{q,r}\,\simeq\,-2\,\mathsf{d}_{2q+1}\,\mathsf{d}_{2r+1}\, \mathsf{M}_{1,1} 
\end{align}
and the asymptotic form (\ref{Co12}). Note that here the dependence on $q$ and $r$ is totally factorized. As a consequence, when we plug this expression into 
(\ref{Co10}) we get
\begin{align}
\label{Co16}
    \frac{\cC_{2q+1}}{\cG^{(0)}_{2q+1}}~{\simeq}~-\mathsf{M}_{1,1}\Big(\mathsf{d}_{2q+1} - \sqrt{\frac{2q+1}{2q-1}} \,\mathsf{d}_{2q-1}\Big)^2
    ~\underset{\lambda \rightarrow \infty}{\sim}~\frac{1}{2}\,\Big(\frac{4\pi}{\sqrt{\lambda}}\,\sqrt{2q+1}\,q\Big)^2+O(\lambda^{-\frac{3}{2}})~.
\end{align}

Let us compare this behavior with that of the 2-point function $\cG_{2q+1}= \vev{O_{2q+1}\, O_{2q+1}}$, for which one has \cite{Billo:2022fnb}
\begin{align}
    \label{Co17}
        \frac{\cG_{2q+1}}{\cG^{(0)}_{2q+1}}  
        ~\underset{\lambda \rightarrow \infty}{\sim}~\frac{8\pi^2}{\lambda} (2q+1) q~.
\end{align}
Taking the ratio between (\ref{Co16}) and (\ref{Co17}), the $\lambda$-dependence drops out and we simply obtain
\begin{align}
    \label{Co18}
        \frac{\cC_{2q+1}}{\cG_{2q+1}} ~\underset{\lambda \rightarrow \infty}{\sim}~ q+O(\lambda^{-\frac{1}{2}})~,  
\end{align}
which is (\ref{ratioE}) for $p=2q+1$.

\section{Conclusions}
\label{sec:conclusions}
Using matrix-model techniques based on supersymmetric localization, we have presented a detailed analysis of the quantities
\begin{align}
    \cC_p=\partial_{\tau_p} \partial_{\,\overbar{\tau}_p}\partial_m^2 \log\cZ_{\mathbf{E}^*}\Big|_{m=0}
\end{align}
which provide the integrated correlators of two Coulomb-branch operators and two moment-map operators in the \textbf{E}-theory. As mentioned in the Introduction, this theory admits sectors with observables that are planar equivalent to those of $\cN=4$ SYM. These include, for example, the coefficients in the 2- and 3-point functions of Coulomb-branch operators with even dimensions \cite{Beccaria:2020hgy,Beccaria:2021hvt,Billo:2022xas}. Here we have shown that also the integrated correlators $\cC_p$ with even $p$ belong to this class. Indeed, comparing (\ref{C2is4}) and (\ref{ppM02}) with $p=2q$, in the planar limit we find
\begin{align}
     \cC_{2q}=\cC_{2q}^{(0)}=\cG_{2q}^{(0)}\,\big(\mathsf{M}_{2q,2q}-2q\,\mathsf{M}_{1,1}\big)~.
     \label{C2qconcl}
\end{align}
On the other hand, there are observables of the \textbf{E}-theory which are not equivalent to those of $\cN=4$ SYM even in the planar limit. Among these there are the 2- and 3-point functions of Coulomb-branch operators with odd dimensions \cite{Beccaria:2020hgy,Beccaria:2021hvt,Billo:2022xas}, but also the integrated correlators $\cC_p$ with odd $p$. This is easily seen in the simplest case $p=3$. In fact, in $\cN=4$ SYM we have (see (\ref{ppM02}))
\begin{align}
    \cC_3^{(0)}=\frac{3N^3}{8}\,\big(\mathsf{M}_{3,3}-3\,\mathsf{M}_{1,1}\big)~,
    \label{C30concl}
\end{align}
whereas in the \textbf{E}-theory we have (see (\ref{C3}))
\begin{align}
    \cC_3=-\frac{3N^3}{8}\,\Big[
    \mathsf{d}_{3}^2\,\mathsf{M}_{1,1} +\big(\mathsf{D}^{\mathrm{odd}} \,\mathsf{M}^{\mathrm{odd}} \,\mathsf{D}^{\mathrm{odd}}\big)_{1,1}\Big]~.
    \label{C3concl}
\end{align}
These two expressions are clearly different. Similar, but more complicated, formulas can be worked out for $p=2q+1$ with the result that
\begin{align}
    \cC_{2q+1}\not= \cC_{2q+1}^{(0)}~.
\end{align}
The matrix-model techniques described in the previous sections allow us to express the integrated correlators $\cC_p$ (but also $\cC_p^{(0)}$) through the coefficients
$\mathsf{M}_{n,m}$ and $(\mathsf{D}^{\mathrm{odd}})_{n,m}$ which in turn are given in terms of convolutions of Bessel functions and are valid for all values of the 't Hooft coupling $\lambda$. Thus, the resulting expressions for the integrated correlators, like those in (\ref{C2qconcl})--(\ref{C3concl}), are exact in the planar limit.

When $\lambda$ is small, we can expand these exact integrated correlators and find their perturbative expansions in the weak-coupling regime. For example, from (\ref{C3concl}) we find
\begin{align}
    \cC_3= \frac{3N^3}{8}\,\Big[\,\frac{9\,
    \zeta_3}{2}\,\hat{\lambda}
    -\frac{45\,\zeta_5}{2}\,\hat{\lambda}^2
    +105\,\zeta_7 \,\hat{\lambda}^3
    -\frac{1890\,\zeta_9+45\,\zeta_3\,\zeta_5}{4}\, \hat{\lambda}^4
    +O(\hat{\lambda}^5)\,\Big]~.
    \label{weakC3}
\end{align}
where $\hat{\lambda}=\lambda/(4\pi^2)$. Doing a similar expansion for $\cC_3^{(0)}$ in (\ref{C30concl}), we get
\begin{align}
    \cC_3^{(0)}&= \frac{3N^3}{8}\,\Big[\,\frac{9\,
    \zeta_3}{2}\,\hat{\lambda}
    -\frac{45\,\zeta_5}{2}\,\hat{\lambda}^2
    +\frac{735\,\zeta_7}{8} \,\hat{\lambda}^3
    -\frac{2835\,\zeta_9}{8}\, \hat{\lambda}^4
    +O(\hat{\lambda}^5)\,\Big]
    \label{weakC30}~.
\end{align}
It is interesting to observe that the difference between the \textbf{E}-theory and $\cN=4$ SYM starts at three loops with the terms proportional to $\zeta_7$. This makes checking such a difference with ordinary field-theory diagrammatic methods quite challenging. Nevertheless, it would be interesting to carry out this 3-loop test.

When $\lambda$ is large, we can take advantage of the asymptotic behavior of the coefficients $\mathsf{M}_{n,m}$ and $(\mathsf{D}^{\mathrm{odd}})_{n,m}$ for $\lambda\to\infty$ that can be deduced from their exact expressions in terms of Bessel functions. Clearly, in the case of even integrated correlators, because of (\ref{C2qconcl}), we obtain the same simple leading behavior as in $\cN=4$ SYM given in (\ref{C2qstrong}). What is less obvious is that a similar simple result also holds for the odd integrated correlators, as shown in (\ref{Co17}). This is a highly non-trivial result since in the \textbf{E}-theory the integrated correlators $\cC_{2q+1}$ and the 2-point functions $\cG_{2q+1}$ are two different functions of the 't Hooft coupling. However, at leading order in the strong-coupling expansions these two functions become identical up to a simple factor related to the conformal dimensions of the Coulomb-branch operators. Therefore, in the planar limit $N\to\infty$ we can conclude that 
\begin{align}
  \lim_{\lambda\to\infty} \, \frac{\cC_p}{\cG_p}= \frac{p-1}{2}~,
  \label{ratioconcl}
\end{align}
for any $p$, just like in $\cN=4$ SYM. 

It would be very interesting to retrieve this simple strong-coupling result with a supergravity calculation using the AdS/CFT correspondence. This analysis would also help to understand the meaning in the holographic dual theory of the integrated insertions of two moment-map operators or, equivalently, of the matrix-model operator $\cM$. Extensions of our present analysis to other $\cN=2$ superconformal gauge theories, like for example the circular quiver theories considered in \cite{Billo:2021rdb,Billo:2022gmq,Billo:2022fnb} which also have a simple holographic dual, would also be interesting and could be useful to clarify the whole picture in a set-up where supersymmetry is not maximal.

\vskip 1cm
\noindent {\large {\bf Acknowledgments}}
\vskip 0.2cm
We would like to thank F. Galvagno and P. Vallarino for many useful discussions.
This research is partially supported by the MUR PRIN contract 2020KR4KN2 ``String Theory as a bridge between Gauge Theories and Quantum Gravity'' and by the INFN project ST\&FI
``String Theory \& Fundamental Interactions''. The work of AP is supported  by the Deutsche Forschungsgemeinschaft (DFG, German Research Foundation) via the Emmy Noether program ``Exploring the landscape of string theory flux vacua using exceptional field theory” (project number 426510644). 

\vskip 1cm

\appendix

\section{The operator \texorpdfstring{$\mathcal{M}$}{} }
\label{appendixM}
Here we give some details on how to derive the large-$N$ expansion of the operator $\cM$ presented in (\ref{M012}). We recall that the initial expression of $\cM$ is (see (\ref{Mis}))
\begin{align}
    \mathcal{M}=-\sum_{n=1}^{\infty}\sum_{\ell=0}^{2n}(-1)^{n}\,\frac{(2n+1)!\,\zeta_{2n+1}}{(2n-\ell)!\,\ell!}\,\Big(\frac{\lambda}{8\pi^2 N}\Big)^{n} \tr a^{2n-\ell}\tr a^{\ell}~,
    \label{Misapp}
\end{align}
and that the $\cP$ operators are defined by (see \ref{avsP}))
\begin{align}
    \tr a^k=\Big(\frac{N}{2}\Big)^{\frac{k}{2}}\,\sum_{\ell=0}^{\lfloor\frac{k-1}{2}\rfloor}
    \sqrt{k-2\ell} \,\binom{k}{\ell} \,\mathcal{P}_{k-2\ell}+
    \langle \tr a^k\rangle_0~.
    \label{avsPapp}
\end{align}
When we insert (\ref{avsPapp}) into (\ref{Misapp}) we find terms with zero, one or two $\cP$ operators given, respectively, by
\begin{subequations}
    \begin{align}
    \mathcal{M}^{(0)}&=-\sum_{n=1}^{\infty}\sum_{\ell=0}^{n}\frac{(-1)^{n}\,(2n+1)!\,\zeta_{2n+1}}{(2n-2\ell)!\,(2\ell)!}\,\Big(\frac{\lambda}{8\pi^2 N}\Big)^{n} \langle \tr a^{2n-2\ell}\rangle_0\,\langle \tr a^{2\ell}\rangle_0~,
    \label{M0isapp}\\[2mm]
    \mathcal{M}^{(1)}&=-2\sum_{n=1}^{\infty}\sum_{\ell=0}^{n}\sum_{q=0}^\ell
    \frac{(-1)^{n}\,(2n+1)!\,\zeta_{2n+1}}{(2n-2\ell)!\,(2\ell)!}\,\Big(\frac{\lambda}{8\pi^2 N}\Big)^{n} \Big(\frac{N}{2}\Big)^\ell
    \sqrt{2\ell-2q}\,\binom{2\ell}{q}\,\cP_{2\ell-2q}\,\big\langle \tr a^{2n-2\ell}\big\rangle_0~,
    \label{M1isapp}\\[-2mm]
    \mathcal{M}^{(2)}&=-\sum_{n=1}^\infty\sum_{\ell=0}^{2n}\sum_{k=0}^{\lfloor\frac{2n-\ell-1}{2}\rfloor}\sum_{q=0}^{\lfloor\frac{\ell-1}{2}\rfloor}
    \frac{(-1)^n\,(2n+1)!\,\zeta_{2n+1}}{(2n-\ell-k)!\,k!\,(\ell-q)!\,q!}\,\Big(\frac{\lambda}{16\pi^2}\Big)^{n}\,\times\notag\\[1mm]
    &\qquad\qquad\qquad\qquad\times\,\sqrt{2n-\ell-2k}\,\sqrt{\ell-2q}\,\cP_{2n-\ell-2k}\,\cP_{\ell-2q}~.\label{M2isapp}
\end{align}
\end{subequations}
Notice that in (\ref{M0isapp}) and (\ref{M1isapp}) we have taken into account that only the traces of even powers of $a$ have non-zero vacuum expectation values.
In the large-$N$ expansion these vacuum expectation values can be obtained using the recursion relations based on the fusion/fission identities discussed in \cite{Billo:2017glv,Beccaria:2020hgy}, with the result that
\begin{align}
    \langle \tr a^{2\ell}\rangle_0=\frac{N^{\ell+1}}{2^\ell}\frac{(2\ell)!}{\ell!\,(\ell+1)!}-\frac{N^{\ell-1}}{2^{\ell+1}}\frac{(2\ell)!}{\ell!\,(\ell-1)!}\Big(1-\frac{\ell-1}{6}\Big)+O(N^{\ell-3})~.
    \label{t2l}
\end{align}
Inserting this into (\ref{M0isapp}), after some straightforward algebra we get
\begin{align}
    \mathcal{M}^{(0)}&=-N^2\sum_{n=1}^{\infty}\sum_{\ell=0}^{n}\frac{(-1)^{n}\,(2n+1)!\,\zeta_{2n+1}}{(n-\ell+1)!\,(n-\ell)!\,\ell!\,(\ell+1)!}\,\Big(\frac{\lambda}{16\pi^2}\Big)^{n}\notag\\
    &\quad+\sum_{n=1}^{\infty}\sum_{\ell=0}^{n}\frac{(-1)^{n}\,(2n+1)!\,\zeta_{2n+1}}{(n-\ell+1)!\,(n-\ell)!\,\ell!\,(\ell-1)!}\,\Big(1-\frac{\ell-1}{6}\Big)\Big(\frac{\lambda}{16\pi^2}\Big)^{n}+O(N^{-2})~.\label{M0isapp1}
\end{align}
Through the identity
\begin{align}
    (2n+1)!\,\zeta_{2n+1}=\int_0^\infty\!\frac{dt}{t}\,\frac{(t/2)^2}{\sinh(t/2)^2}\,t^{2n}~,
    \label{idzeta}
\end{align}
we rewrite (\ref{M0isapp1}) as
\begin{align}
    \mathcal{M}^{(0)}&=-N^2\int_0^\infty\!\frac{dt}{t}\,\frac{(t/2)^2}{\sinh(t/2)^2}\,\sum_{n=1}^{\infty}\sum_{\ell=0}^{n}\frac{(-1)^{n}}{(n-\ell+1)!\,(n-\ell)!\,\ell!\,(\ell+1)!}\,\Big(\frac{t\sqrt{\lambda}}{4\pi}\Big)^{2n}\notag\\
    &\quad+\int_0^\infty\!\frac{dt}{t}\,\frac{(t/2)^2}{\sinh(t/2)^2}\,\sum_{n=1}^{\infty}\sum_{\ell=0}^{n}\frac{(-1)^{n}}{(n-\ell+1)!\,(n-\ell)!\,\ell!\,(\ell-1)!}\,\Big(\frac{t\sqrt{\lambda}}{4\pi}\Big)^{2n}\label{M0isapp2}\\
    &\quad-\frac{1}{6}\int_0^\infty\!\frac{dt}{t}\,\frac{(t/2)^2}{\sinh(t/2)^2}\,\sum_{n=1}^{\infty}\sum_{\ell=0}^{n}\frac{(-1)^{n}}{(n-\ell+1)!\,(n-\ell)!\,\ell!\,(\ell-2)!}\,\Big(\frac{t\sqrt{\lambda}}{4\pi}\Big)^{2n}+O(N^{-2})~.\notag
\end{align}
After relabeling the summation indices we can resum the above series using the expansion of the Bessel functions of the first kind
\begin{align}
    J_\alpha(x)=\sum_{n=0}^\infty\frac{(-1)^n}{n!\,(n+\alpha)!}\,\Big(\frac{x}{2}\Big)^{2n+\alpha}~,
\end{align}
and get
\begin{align}
    \mathcal{M}^{(0)}&=N^2\int_0^\infty\!\frac{dt}{t}\,\frac{(t/2)^2}{\sinh(t/2)^2} \bigg[1-\frac{16\pi^2}{t^2\lambda}\,J_1\Big(\frac{t\sqrt{\lambda}}{2\pi}\Big)^2\bigg]\notag\\
    &\quad-\int_0^\infty\!\frac{dt}{t}\,\frac{(t/2)^2}{\sinh(t/2)^2}
    \,J_1\Big(\frac{t\sqrt{\lambda}}{2\pi}\Big)\,J_1\Big(\frac{t\sqrt{\lambda}}{2\pi}\Big)\label{M0isapp3}
    \\
    &\quad
    -\frac{1}{12}\int_0^\infty \!\frac{dt}{t}\,\frac{(t/2)^2}{\sinh(t/2)^2} \,\Big(\frac{t \sqrt{\lambda }}{2 \pi }\Big)
    J_1\Big(\frac{t \sqrt{\lambda }}{2 \pi }\Big)\, J_2\Big(\frac{t \sqrt{\lambda }}{2   \pi }\Big)+O(N^{-2})~.\notag
\end{align}
In terms of the matrix elements defined in (\ref{Mmatrix}) we have
\begin{align}
    \mathcal{M}^{(0)}=N^2\,\mathsf{M}_{0,0}+\mathsf{M}_{1,1}-\frac{1}{6}\,\sum_{k=1}^\infty\sqrt{2k+1}\,\mathsf{M}_{1,2k+1}+O(N^{-2})
\end{align}
where we have used the identity (\ref{M12k+1}) to rewrite the last term. This is the expression reported in (\ref{M0}) of the main text.

With very similar manipulations we can treat the term $\cM^{(1)}$ in (\ref{M1isapp}) which, after using (\ref{t2l}), becomes
\begin{align}
    \mathcal{M}^{(1)} &=-2N\sum_{n=1}^{\infty}\sum_{\ell=0}^{n}\sum_{q=0}^\ell
    \frac{(-1)^{n}\,(2n+1)!\,\zeta_{2n+1}}{(2\ell-q)!\,q!\,(n-\ell+1)!\,(n-\ell)!}\,\Big(\frac{\lambda}{16\pi^2}\Big)^{n}
    \sqrt{2\ell-2q}\,\cP_{2\ell-2q}+O(N^{-1})~.
    \label{M1isapp1}
\end{align}
Proceeding as before, we find
\begin{align}
    \mathcal{M}^{(1)} &=-2N\!\!\int_0^\infty\!\frac{dt}{t}\,\frac{(t/2)^2}{\sinh(t/2)^2}\,\sum_{k=1}^\infty\sum_{q=0}^{\infty}\sum_{p=0}^{\infty}
    \frac{(-1)^{k+p+q}\,\sqrt{2k}\,\cP_{2k}}{p!\,(p+1)!\,(q+2k)!\,q!}\,\Big(\frac{t\sqrt{\lambda}}{4\pi}\Big)^{2p+2q+2k}+O(N^{-1})\notag\\
    &=-2N\!\!\int_0^\infty\!\frac{dt}{t}\,\frac{(t/2)^2}{\sinh(t/2)^2}\,\sum_{k=1}^\infty(-1)^k\,\sqrt{2k}\,\cP_{2k}\,\Big(\frac{4\pi }{t \sqrt{\lambda }}\Big)\,J_1\Big(\frac{t \sqrt{\lambda }}{2\pi }\Big)\,J_{2k}\Big(\frac{t \sqrt{\lambda }}{2\pi }\Big)
    +O(N^{-1})
    \notag\\
    &=2N\sum_{k=1}^\infty \mathsf{M}_{0,2k}\,\cP_{2k}+O(N^{-1})
   \label{M1isapp2}
\end{align}
which is the expression reported in (\ref{M1}) of the main text.

Finally, using the identity (\ref{idzeta}) in (\ref{M2isapp}) and relabeling the summation indices we get
\begin{align}
    \mathcal{M}^{(2)}&=-\int_0^\infty\!\frac{dt}{t}\,\frac{(t/2)^2}{\sinh(t/2)^2}\,\sum_{p=1}^\infty\sum_{k=0}^{\infty}\frac{(-1)^{\frac{p}{2}+k}\,\sqrt{p}}{k!\,(k+p)!}\,\Big(\frac{t\sqrt{\lambda}}{4\pi}\Big)^{p+2k}\,\cP_{p}\,\times\notag\\
    &\qquad\qquad\qquad~~\times\sum_{q=1}^{\infty}\sum_{\ell=0}^{\infty}\
    \frac{(-1)^{\frac{q}{2}+\ell}\,\sqrt{q}}{\ell!\,(\ell+q)!}\,\Big(\frac{t\sqrt{\lambda}}{4\pi}\Big)^{q+2\ell}\,\cP_{q}\notag\\
    &=-\int_0^\infty\!\frac{dt}{t}\,\frac{(t/2)^2}{\sinh(t/2)^2}\,\sum_{p,q=1}^\infty(-1)^{\frac{p+q}{2}}\,\sqrt{p\,q}\,J_p\Big(\frac{t\sqrt{\lambda}}{2\pi}\Big)\,J_p\Big(\frac{t\sqrt{\lambda}}{2\pi}\Big)\,\cP_p\,\cP_q
    \notag\\&
    =\sum_{p,q=1}^\infty(-1)^{pq}\,\mathsf{M}_{p,q}\,\cP_p\,\cP_q~.\label{M2isapp1}
\end{align}
Taking into account that $\cP_1=0$ due to the tracelessness of $a$, we easily recognize the same expression reported in (\ref{M2}) of the main text.

\section{Sum rules for \texorpdfstring{$\mathsf{M}_{n,m}$}{}}
\label{app:sumrules}
Here we prove some useful relations satisfied by the matrix elements $\mathsf{M}_{n,m}$ or combinations thereof.

\paragraph{$\bullet$} Proof of
\begin{align}
    \sum_{k=1}^\infty \sqrt{2k}\,\mathsf{M}_{0,2k}=-\mathsf{M}_{1,1}
    \label{M02kM11}
\end{align}
Using (\ref{M0n}), we can rewrite the left-hand side as
\begin{align}
    \sum_{k=1}^\infty \sqrt{2k}\,\mathsf{M}_{0,2k}&=-\int_0^\infty\!\frac{dt}{t}\,\frac{(t/2)^2}{\sinh(t/2)^2}\,J_1\Big(\frac{t \sqrt{\lambda }}{2\pi }\Big) \sum_{k=1}^\infty(-1)^k\, \frac{4k\,J_{2k}\big(\frac{t \sqrt{\lambda }}{2\pi }\big)}{\big(\frac{t \sqrt{\lambda }}{2\pi }\big)}\notag\\
    &=-\int_0^\infty\!\frac{dt}{t}\,\frac{(t/2)^2}{\sinh(t/2)^2}\,J_1\Big(\frac{t \sqrt{\lambda }}{2\pi }\Big) \sum_{k=1}^\infty(-1)^k\, \bigg[J_{2k-1}\Big(\frac{t \sqrt{\lambda }}{2\pi }\Big)+
    J_{2k+1}\Big(\frac{t \sqrt{\lambda }}{2\pi }\Big)\bigg]
\end{align}
where the last line follows from the recursion relation of the Bessel functions:
\begin{align}
   \frac{2\alpha\,J_\alpha(x)}{x}=J_{\alpha-1}(x)+J_{\alpha+1}(x)~.
    \label{recursion}
\end{align}
Performing the sum over $k$ we find
\begin{align}
    \sum_{k=1}^\infty \sqrt{2k}\,\mathsf{M}_{0,2k}&=\int_0^\infty\!\frac{dt}{t}\,\frac{(t/2)^2}{\sinh(t/2)^2}\,J_1\Big(\frac{t \sqrt{\lambda }}{2\pi }\Big)J_1\Big(\frac{t \sqrt{\lambda }}{2\pi }\Big) =-\mathsf{M}_{1,1}~.
    \label{proof1}
\end{align}

\paragraph{$\bullet$} Proof of
\begin{align}
    \sum_{k=1}^\infty\sqrt{2k+1}\,\mathsf{M}_{1,2k+1}=\frac{1}{2}\int_0^\infty \!\frac{dt}{t}\,\frac{(t/2)^2}{\sinh(t/2)^2} \,\Big(\frac{t \sqrt{\lambda }}{2 \pi }\Big)
    J_1\Big(\frac{t \sqrt{\lambda }}{2 \pi }\Big)\, J_2\Big(\frac{t \sqrt{\lambda }}{2   \pi }\Big)
    \label{M12k+1}
\end{align}
Using (\ref{Mnm}) we rewrite the left-hand side as
\begin{align}
    \sum_{k=1}^\infty\sqrt{2k+1}\,\mathsf{M}_{1,2k+1}&=-\int_0^\infty\!\frac{dt}{t}\,\frac{(t/2)^2}{\sinh(t/2)^2}\,\Big(\frac{t \sqrt{\lambda }}{2\pi }\Big) J_1\Big(\frac{t \sqrt{\lambda }}{2\pi }\Big)\sum_{k=1}^\infty(-1)^k\, \frac{(2k+1)\,J_{2k+1}\big(\frac{t \sqrt{\lambda }}{2\pi }\big)}{\big(\frac{t \sqrt{\lambda }}{2\pi }\big)}\notag\\
    &=\frac{1}{2}\int_0^\infty \!\frac{dt}{t}\,\frac{(t/2)^2}{\sinh(t/2)^2} \,\Big(\frac{t \sqrt{\lambda }}{2 \pi }\Big)
    J_1\Big(\frac{t \sqrt{\lambda }}{2 \pi }\Big)\, J_2\Big(\frac{t \sqrt{\lambda }}{2   \pi }\Big)
\end{align}
where the sum over $k$ has been performed using again the recursion relation (\ref{recursion}) of the Bessel functions.

\paragraph{$\bullet$} Proof of
\begin{align}
    \Tr \mathsf{M}^{\mathrm{odd}}=\frac{1}{4}\int_0^\infty\!\frac{dt}{t}\,\frac{(t/2)^2}{\sinh(t/2)^2}\,
    \Big(\frac{t \sqrt{\lambda }}{2 \pi }\Big)^2\,
     \bigg[\,J_1\Big(\frac{t \sqrt{\lambda }}{2\pi }\Big)\, J_3\Big(\frac{t \sqrt{\lambda }}{2\pi }\Big)-J_2\Big(\frac{t
   \sqrt{\lambda }}{2\pi }\Big)^2\,\bigg] 
   \label{TrModd}
\end{align}
Using (\ref{Mnm}) the left-hand side becomes
\begin{align}
    \Tr \mathsf{M}^{\mathrm{odd}}&=-\int_0^\infty\!\frac{dt}{t}\,\frac{(t/2)^2}{\sinh(t/2)^2}\,
    \sum_{k=1}^\infty (2k+1)\,J_{2k+1}\Big(\frac{t \sqrt{\lambda }}{2\pi }\Big)\,
    J_{2k+1}\Big(\frac{t \sqrt{\lambda }}{2\pi }\Big)\notag\\
    &=-\int_0^\infty\!\frac{dt}{t}\,\frac{(t/2)^2}{\sinh(t/2)^2}\,
    \cK\Big(\frac{t \sqrt{\lambda }}{2\pi },\frac{t \sqrt{\lambda }}{2\pi }\Big)
       \label{TrModd1}
\end{align}
where 
\begin{align}
    \cK(x,y)=\sum_{k=1}^\infty (2k+1)\,J_{2k+1}(x)\,J_{2k+1}(y)=\frac{x\,y\,\big[x\,J_3(x)\,J_2(y)-y\,J_2(x)\,J_3(y)\big]}{2(x^2-y^2)}
    \label{BesselKernel}
\end{align}
is a Bessel kernel \cite{Tracy:1993xj}. Using the recursion relation (\ref{recursion}) of the Bessel functions and the property
\begin{align}
    2\,J_\alpha^\prime(x)=J_{\alpha-1}(x)-J_{\alpha+1}(x)~,
    \label{derivativeJ}
\end{align}
it is not difficult to show that
\begin{align}
    \cK(x,x)=\frac{x}{4}\,\big[x\,J_2(x)\,J_3^\prime(x)-x\,J_3(x)\,J_2^\prime(x)+J_2(x)\,J_3(x) \big]=-\frac{x^2}{4}\,\big[J_1(x)\,J_3(x)-J_2(x)^2\big]~.
    \label{BesselKernel1}
\end{align}
Upon inserting this result in (\ref{TrModd1}), the identity (\ref{TrModd}) immediately follows.

\paragraph{$\bullet$} Proof of
\begin{align}
    \Tr \mathsf{M}^{\mathrm{even}}=\mathsf{M}_{1,1}+\Tr \mathsf{M}^{\mathrm{odd}}+\frac{1}{2}\int_0^\infty\!\frac{dt}{t}\,\frac{(t/2)^2}{\sinh(t/2)^2}\,
    \Big(\frac{t \sqrt{\lambda }}{2 \pi }\Big)\,
     J_0\Big(\frac{t \sqrt{\lambda }}{2\pi }\Big)\, J_1\Big(\frac{t \sqrt{\lambda }}{2\pi }\Big)
     \label{TrMeven}
\end{align}
Using (\ref{Mnm}) we write the left-hand side as
\begin{align}
    \Tr \mathsf{M}^{\mathrm{even}}&=-\int_0^\infty\!\frac{dt}{t}\,\frac{(t/2)^2}{\sinh(t/2)^2}\,
    \sum_{k=1}^\infty 2k\,J_{2k}\Big(\frac{t \sqrt{\lambda }}{2\pi }\Big)\,
    J_{2k}\Big(\frac{t \sqrt{\lambda }}{2\pi }\Big)\notag\\
    &=-\int_0^\infty\!\frac{dt}{t}\,\frac{(t/2)^2}{\sinh(t/2)^2}\,
    \cH\Big(\frac{t \sqrt{\lambda }}{2\pi },\frac{t \sqrt{\lambda }}{2\pi }\Big)
     \label{TrMeven1}
\end{align}
where $\cH$ is the following Bessel kernel
\begin{align}
    \cH(x,y)=\sum_{k=1}^\infty 2k\,J_{2k}(x)\,J_{2k}(y)&=\frac{x\,y\,\big[x\,J_2(x)\,J_1(y)-y\,J_1(x)\,J_2(y)\big]}{2(x^2-y^2)}\notag\\
    &=\frac{x\,y\,\big[y\,J_0(y)\,J_1(x)-x\,J_0(x)\,J_1(y)\big]}{2(x^2-y^2)}~.
    \label{BesselKernelH}
\end{align}
From the last expression we easily obtain
\begin{align}
    \cH(x,x)=\frac{x}{4}\,\big[x\,J_0(x)\,J_1^\prime(x)-x\,J_1(x)\,J_0^\prime(x)-J_0(x)\,J_1(x) \big]~.
    \label{BesselKernelH1}
\end{align}
On the other hand, we have
\begin{align}
    \mathsf{M}_{1,1}+\Tr \mathsf{M}^{\mathrm{odd}}&=\sum_{k=0}^\infty
    M_{2k+1,2k+1}=-\int_0^\infty\!\frac{dt}{t}\,\frac{(t/2)^2}{\sinh(t/2)^2}\,
    \sum_{k=0}^\infty (2k+1)\,J_{2k+1}\Big(\frac{t \sqrt{\lambda }}{2\pi }\Big)\,
    J_{2k+1}\Big(\frac{t \sqrt{\lambda }}{2\pi }\Big)\notag\\
    &=-\int_0^\infty\!\frac{dt}{t}\,\frac{(t/2)^2}{\sinh(t/2)^2}\,
    \cG\Big(\frac{t \sqrt{\lambda }}{2\pi },\frac{t \sqrt{\lambda }}{2\pi }\Big)
    \label{trModd0}
\end{align}
where now the Bessel kernel is
\begin{align}
    \cG(x,y)=\sum_{k=0}^\infty (2k+1)\,J_{2k+1}(x)\,J_{2k+1}(y)=\frac{x\,y\,\big[x\,J_1(x)\,J_0(y)-y\,J_1(y)\,J_0(x)\big]}{2(x^2-y^2)}~.
    \label{BesselKernelG}
\end{align}
From the properties of the Bessel functions we deduce that
\begin{align}
    \cG(x,x)=\frac{x}{4}\,\big[x\,J_0(x)\,J_1^\prime(x)-x\,J_1(x)\,J_0^\prime(x)+J_0(x)\,J_1(x) \big]=\cH(x,x)+\frac{x}{2}\,J_0(x)\,J_1(x)~.
\end{align}
Using this relation in (\ref{trModd0}) and comparing with (\ref{TrMeven1}), it is straightforward to obtain the identity (\ref{TrMeven}).

It is interesting to notice that these same identities follow from topological recursion formulas as discussed in detail in \cite{Chester:2019pvm}.

\section{The calculation of \texorpdfstring{$\langle \cP_k\rangle$}{} }
\label{appendixPka}
The 1-point function of the operator $\cP_k$ in the matrix model of the \textbf{E}-theory is
\begin{align}
    \langle \cP_k\rangle=\frac{\displaystyle{\big\langle \cP_k\,\rme^{-S_0}\big\rangle_0}\phantom{\Big|}}{\displaystyle{\big\langle\,\rme^{-S_0}\big\rangle_0}\phantom{\Big|}}~.
    \label{Pkapp}
\end{align}
Since the interaction action $S_0$ contains only odd operators (see (\ref{S0P})) and the vacuum expectation values of products of only odd operators vanish, the 1-point function of an odd operator is trivially zero, namely
\begin{align}
    \langle \cP_{2k+1}\rangle=0~.
    \label{Poddapp}
\end{align}
Instead, the 1-point function $\langle \cP_{2k}\rangle$ does not vanish. Expanding (\ref{Pkapp}) in powers of $S_0$ and taking into account that
$\langle \cP_{2k}\rangle_0=0$, we have
\begin{align}
    \langle \cP_{2k}\rangle&=
    -\langle \cP_{2k}\,S_0\rangle_0+
    \frac{1}{2}\,\langle \cP_{2k}\,S_0\,S_0\rangle_0
    -\langle \cP_{2k}\,S_0\rangle_0\,\langle S_0\rangle_0+\ldots\notag\\
    &=\frac{1}{2}\sum_{n,m=1}^{\infty}
    \big\langle\cP_{2k}\,\cP_{2n+1}\,\cP_{2m+1}\big\rangle_0\,\mathsf{X}_{n,m}^{\mathrm{odd}} \label{P2kapp}\\
    &~~+\frac{1}{8}\sum_{n,m,p,q=1}^{\infty}
    \big\langle\cP_{2k}\,\cP_{2n+1}\,\cP_{2m+1}\,\cP_{2p+1}\,\cP_{2q+1}\big\rangle_0\,\mathsf{X}_{n,m}^{\mathrm{odd}}
    \,\,\mathsf{X}_{p,q}^{\mathrm{odd}}\notag\\
    &~~-\frac{1}{4}\sum_{n,m,p,q=1}^{\infty}
    \big\langle\cP_{2k}\,\cP_{2n+1}\,\cP_{2m+1}\big\rangle_0\,\big\langle\cP_{2p+1}\,\cP_{2q+1}\big\rangle_0\,\mathsf{X}_{n,m}^{\mathrm{odd}}
    \,\,\mathsf{X}_{p,q}^{\mathrm{odd}}+\ldots\notag
\end{align}
where the second equality follows from the use of the expression of $S_0$ given in (\ref{S0P}). Exploiting the form (\ref{PPPodd}) of the 3-point functions and applying Wick's theorem with the free contraction (\ref{PkPl0}), we can simplify (\ref{P2kapp}) and get
\begin{align}
    \langle \cP_{2k}\rangle&=\frac{1}{2N}
\sum_{n,m=1}^\infty d_{2k,2n+1,2m+1}\,
\big(\mathsf{X}^{\mathrm{odd}}+(\mathsf{X}^{\mathrm{odd}})^2+\ldots \big)_{n,m}
\notag\\
&=\frac{\sqrt{2k}}{2N}\sum_{n,m=1}^\infty \sqrt{(2n+1)\,(2m+1)}
\,\big(\mathsf{D}^{\mathrm{odd}}-\bone\big)_{n,m}~.
\label{P2kapp1}
\end{align}
In the second line we have used the expression of the coefficients $d_{2k,2n+1,2m+1}$ given in (\ref{dkln}) and introduced the odd ``propagator'' 
$\mathsf{D}^{\mathrm{odd}}$ as in (\ref{Dnm}).

Since the dependence on $k$ is entirely in the prefactor, we can rewrite (\ref{P2kapp1}) as
\begin{align}
    \langle \cP_{2k}\rangle=\sqrt{k}\,\langle \cP_{2}\rangle~.
    \label{P2kvsP2}
\end{align}
On the other hand, from (\ref{avsP}) it easily follows that
\begin{align}
    \langle\cP_2\rangle=\frac{\sqrt{2}}{N}\big(\langle\tr a^2\rangle
    -\langle \tr a^2\rangle_0\big)
    =\frac{\sqrt{2}}{N}\Big(\langle\tr a^2\rangle-\frac{N^2-1}{2}\Big)~.
    \label{P2E}
\end{align}
In turn, $\langle\tr a^2\rangle$ can be directly computed from the partition function
\begin{align}
    \cZ_{\mathbf{E}}=\int\!da~\rme^{-\tr a^2-S^{(0)}_{\mathrm{int}}}=\Big(\frac{8\pi^2N}{\lambda}\Big)^{\frac{N^2-1}{2}}\int\!
    d\tilde{a}~\rme^{-\frac{8\pi^2\,N}{\lambda}\,\tr \tilde{a}^2-\tilde{S}^{(0)}_{\mathrm{int}}}
\end{align}
where in the last step we have performed the inverse rescaling (\ref{rescaling}) to remove all dependence on $\lambda$ from the interaction action and produce the prefactor in front of the quadratic term. In fact, we have
\begin{align}
    \lambda\,\partial_\lambda\log\cZ_{\mathbf{E}}=\langle\tr a^2\rangle
    -\frac{N^2-1}{2}~.
\end{align}
Using this result in (\ref{P2E}), we get
\begin{align}
    \langle\cP_2\rangle=\frac{\sqrt{2}}{N}\,
    \lambda\,\partial_\lambda\log\cZ_{\mathbf{E}}=-\frac{\sqrt{2}}{N}\,
    \lambda\,\partial_\lambda\cF_{\mathbf{E}}
    \label{P2E1}
\end{align}
where $\cF_{\mathbf{E}}$ is the free energy. Finally, from (\ref{P2kvsP2}) we can conclude that
\begin{align}
    \langle\cP_{2k}\rangle=-\frac{\sqrt{2k}}{N}\,
    \lambda\,\partial_\lambda\cF_{\mathbf{E}}~,
\end{align}
as reported in (\ref{vevP2k}) of the main text.

\section{Strong-coupling methods}
\label{app:strong}
In this appendix we provide a few technical details on the derivation of the strong-coupling behavior of some of the quantities considered in the main text.
First of all, we observe that by setting
\begin{align}
    t=\sqrt{x}\quad\mbox{and}\quad g=\frac{\sqrt{\lambda}}{2\pi}~,
    \label{tglambda}
\end{align}
the matrix elements $\mathsf{X}_{n,m}$ and $\mathsf{M}_{n,m}$ defined in (\ref{Xij}) and (\ref{Mnm}) can be conveniently rewritten as
\begin{subequations}
    \begin{align}
    \mathsf{X}_{n,m}&=(-1)^{\frac{n+m+2nm}{2}}\,\sqrt{n\,m}\int_0^\infty\!\frac{dx}{x}\,\chi(\sqrt{x})\,J_n(g\sqrt{x})\,J_m(g\sqrt{x})~,\label{Xnmchi}\\[1mm]
    \mathsf{M}_{n,m}&=(-1)^{\frac{n+m+2nm}{2}}\frac{\sqrt{n\,m}}{8}\int_0^\infty\!dx\,\chi(\sqrt{x})\,J_n(g\sqrt{x})\,J_m(g\sqrt{x})~,\label{Mnmchi}
\end{align}%
\label{XMchi}
\end{subequations}
with
\begin{align}
    \chi(x)=-\frac{1}{\sinh(x/2)^2}~.
    \label{chiis}
\end{align}
These formulas are particularly useful for the strong-coupling analysis that we are going to present.

\subsection{Strong-coupling behavior of \texorpdfstring{$-\Tr \big[\mathsf{M}^{\mathrm{odd}}\,(\mathsf{D}^{\mathrm{odd}}-\bone)\big]$}{}}
Here we derive the leading behavior of
\begin{align}
    -\Tr \Big[\mathsf{M}^{\mathrm{odd}}\,(\mathsf{D}^{\mathrm{odd}}-\bone)\Big]=-\sum_{k=1}^{\infty}\Tr \Big[\mathsf{M}^{\mathrm{odd}}\,\big(\mathsf{X}^{\mathrm{odd}}\big)^k\Big]
    \label{Q}
\end{align}
at strong coupling. Let us first consider the term with $k=1$ in the above sum. Using (\ref{XMchi}) we can write it as follows
\begin{align}
    -\Tr \Big[\mathsf{M}^{\mathrm{odd}}\,\mathsf{X}^{\mathrm{odd}}\Big]&=-\frac{1}{8}\int_0^\infty\!\!
    dx\!\int_0^\infty\!\frac{dy}{y}\,\chi(\sqrt{x})\,\chi(\sqrt{y})
    \Big[\sum_{\ell=1}^\infty (2\ell+1)\,J_{2\ell+1}(g\sqrt{x})\,J_{2\ell+1}(g\sqrt{y})\Big]^2\notag\\
    &=-\frac{g^4}{8}\int_0^\infty\!\!
    dx\!\int_0^\infty\!dy\,\,x\,\chi(\sqrt{x})\,\chi(\sqrt{y})\,
    K(g^2 x,g^2 y)^2
    \label{Q1}
\end{align}
where in the second line we have introduced the rescaled Bessel kernel (see (\ref{BesselKernel}))
\begin{align}
    K(t,t')=\frac{\cK(\sqrt{t},\sqrt{t'})}{\sqrt{t\,t'}}=\frac{\sqrt{t}\,J_3(\sqrt{t})\,J_2(\sqrt{t'})-\sqrt{t'}\,J_2(\sqrt{t})\,J_3(\sqrt{t'})}{2(t-t')}
    ~.\label{K}
\end{align}
At leading order when $g\to\infty$ we can replace the Bessel kernel in (\ref{Q1}) with its asymptotic behavior when the arguments are large \cite{Belitsky:2020qrm,Belitsky:2020qir}, namely
\begin{align}
    K(g^2 x,g^2 y)
    ~\underset{g \rightarrow \infty}{\sim}~
    \frac{1}{2\pi g (x\,y)^{1/4}}\Big[\,\frac{\sin (g\sqrt{x}-g\sqrt{y})}{(g\sqrt{x}-g\sqrt{y})}
    -\frac{\cos(g\sqrt{x}+g\sqrt{y})}{g\sqrt{x}+g\sqrt{y}}\Big]~.
    \label{Kexp}
\end{align}
The cosine part is a rapidly oscillating function which does not contribute to the integrals and can be discarded. The sine part, instead, is peaked around $x=y$ and, when we apply it to an arbitrary test function $f$, we get\,%
\footnote{See also eq. (4.5) of \cite{Belitsky:2020qrm}.}
\begin{align}
    \int_0^\infty\!dy\, K(g^2 x,g^2 y)\,f(y) &~\underset{g \rightarrow \infty}{\sim}\int_0^\infty\!dy\,\frac{1}{2\pi g (x\,y)^{1/4}}\,\frac{\sin (g\sqrt{x}-g\sqrt{y})}{(g\sqrt{x}-g\sqrt{y})}\,f(y)\notag\\[3mm]
    &~~~=\frac{1}{g^2}\int_{-\infty}^{+\infty}\!dz\,\frac{\sin z}{\pi z}\,\bigg[\frac{y^{1/4}}{x^{1/4}}\,f(y)\bigg]_{y=(\sqrt{x}+\frac{z}{g})^2}~=~\frac{1}{g^2}\,f(x)
\end{align}
where the last step follows from the Jordan lemma which instructs us to compute the quantity in square brackets at $z=0$. Applying this identity to the $y$ integral in (\ref{Q1}) yields
\begin{align}
    -\Tr \Big[\mathsf{M}^{\mathrm{odd}}\,\mathsf{X}^{\mathrm{odd}}\Big]\simeq-\frac{g^2}{8}\int_0^\infty\!\!
    dx\,\,x\,\chi(\sqrt{x})^2\,K(g^2 x,g^2 x)
    \label{Q1a}
\end{align}
In a very similar way one can show that
\begin{align}
     -\Tr \Big[\mathsf{M}^{\mathrm{odd}}\,(\mathsf{X}^{\mathrm{odd}})^k\Big]\simeq-\frac{g^2}{8}\int_0^\infty\!\!
    dx\,\,x\,\chi(\sqrt{x})^{k+1}\,K(g^2 x,g^2 x)~,
    \label{Qk}
\end{align}
so that (\ref{Q}) becomes
\begin{align}
    -\Tr \Big[\mathsf{M}^{\mathrm{odd}}\,(\mathsf{D}^{\mathrm{odd}}-\bone)\Big]\simeq
    -\frac{g^2}{8}\int_0^\infty\!\!
    dx\,\,x\,\frac{\chi(\sqrt{x})^2}{1-\chi(\sqrt{x})}\,K(g^2 x,g^2 x)~.
    \label{Qstrong}
\end{align}
From (\ref{Kexp}) we see that $K(g^2 x,g^2 x)\simeq\frac{1}{2\pi g \sqrt{x}}$. Inserting this expression into (\ref{Qstrong})
and changing integration variable by setting $x=z^2$, we finally get
\begin{align}
    -\Tr \Big[\mathsf{M}^{\mathrm{odd}}\,(\mathsf{D}^{\mathrm{odd}}-\bone)\Big]\simeq
    -\frac{g}{8\pi}\int_0^\infty\!\!dz\,z^2\,\frac{\chi(z)^2}{1-\chi(z)}=-\frac{g}{8\pi}\Big(\frac{2\pi^2}{3}\Big)=-\frac{\sqrt{\lambda}}{24}~,
    \label{Qstrongfin}
\end{align}
thus proving the claim (\ref{3rtermstrong}). We have also performed a numerical analysis of this expression by computing the perturbative expansion of the left-hand side up to very high orders and then
using the conformal Pad\'e resummation to estimate the behavior of the function for very large values of 
$\lambda$. These numerical results are in full agreement with the analytic derivation presented above.

\subsection{Strong-coupling behavior of \texorpdfstring{$\big(\mathsf{D}^{\mathrm{odd}} \,\mathsf{M}^{\mathrm{odd}} \,\mathsf{D}^{\mathrm{odd}}\big)_{q,r}$}{}}
\label{app:strongDMD}
Here we show that $\big(\mathsf{D}^{\mathrm{odd}} \,\mathsf{M}^{\mathrm{odd}} \,\mathsf{D}^{\mathrm{odd}}\big)_{q,r}$ behaves as $\lambda^{-\frac{3}{2}}$ at strong coupling.
Using the asymptotic expansion (\ref{M11strong}) for the matrix elements of $\mathsf{M}$ we have
\begin{align}
    \big(\mathsf{D}^{\mathrm{odd}} \,\mathsf{M}^{\mathrm{odd}} \,\mathsf{D}^{\mathrm{odd}}\big)_{q,r}&=
    \sum_{\ell,p=1}^\infty
    (\mathsf{D}^{\mathrm{odd}})_{q,\ell}\,\mathsf{M}_{2\ell+1,2p+1} \,(\mathsf{D}^{\mathrm{odd}})_{p,r}\label{DMD1}\\
    &\hspace{-5pt}\underset{\lambda \rightarrow \infty}
    {\sim}-\frac{1}{2}\sum_{\ell=1}^\infty
    (\mathsf{D}^{\mathrm{odd}})_{q,\ell}\,(\mathsf{D}^{\mathrm{odd}})_{\ell,r}+
    \frac{1}{\sqrt{\lambda}}\sum_{\ell,p=1}^\infty \sqrt{\ell\,p}\,(\mathsf{D}^{\mathrm{odd}})_{q,\ell}\,(\mathsf{D}^{\mathrm{odd}})_{p,r}+O(\lambda^{-\frac{3}{2}})~.\notag
\end{align}
Introducing the quantities $\mathsf{d}_q$ defined in (\ref{dis}), we can rewrite the previous expression as
\begin{align}
    \big(\mathsf{D}^{\mathrm{odd}} \,\mathsf{M}^{\mathrm{odd}} \,\mathsf{D}^{\mathrm{odd}}\big)_{q,r}\,\underset{\lambda \rightarrow \infty}
    {\sim}-\frac{1}{2}\sum_{\ell=1}^\infty
    \mathsf{D}^{\mathrm{odd}}_{q,\ell}\,\mathsf{D}^{\mathrm{odd}}_{\ell,r}+
    \frac{1}{\sqrt{\lambda}}\,\mathsf{d}_q\,\mathsf{d}_r+O(\lambda^{-\frac{3}{2}})~.
    \label{DMD2}
\end{align}
Since at strong coupling $\mathsf{d}_q$ is $O(\lambda^{-\frac{1}{2}})$ as we have shown in \cite{Billo:2022fnb} and recalled in (\ref{Co11}), we see that the second term in (\ref{DMD2}) is $O(\lambda^{-\frac{3}{2}})$ when $\lambda\to\infty$. Thus, we are left with investigating the strong-coupling behavior of the first term, namely of
\begin{align}
    \Phi_{q,r}=\sum_{\ell=1}^\infty
    \mathsf{D}^{\mathrm{odd}}_{q,\ell}\,\mathsf{D}^{\mathrm{odd}}_{\ell,r}~.
\label{Phi}
\end{align}
Given that the entries of $\mathsf{D}^{\mathrm{odd}}$ scale as $\lambda^{-1}$, one might be tempted to conclude that $\Phi_{q,r}$ behaves like $\lambda^{-2}$ at strong coupling. However, this conclusion has to be refined, since the sum over $\ell$ in (\ref{Phi}) produces a divergence. A very similar phenomenon occurs in the quantities $\mathsf{d}_q$. Being linear in $\mathsf{D}^{\mathrm{odd}}$, one might think that they should scale as $\lambda^{-1}$, but again a divergence is produced and their correct strong-coupling behavior is actually $\lambda^{-\frac{1}{2}}$
(see (\ref{Co11})). 

\begin{figure}[ht]
    \centering
    \includegraphics[scale=0.38]{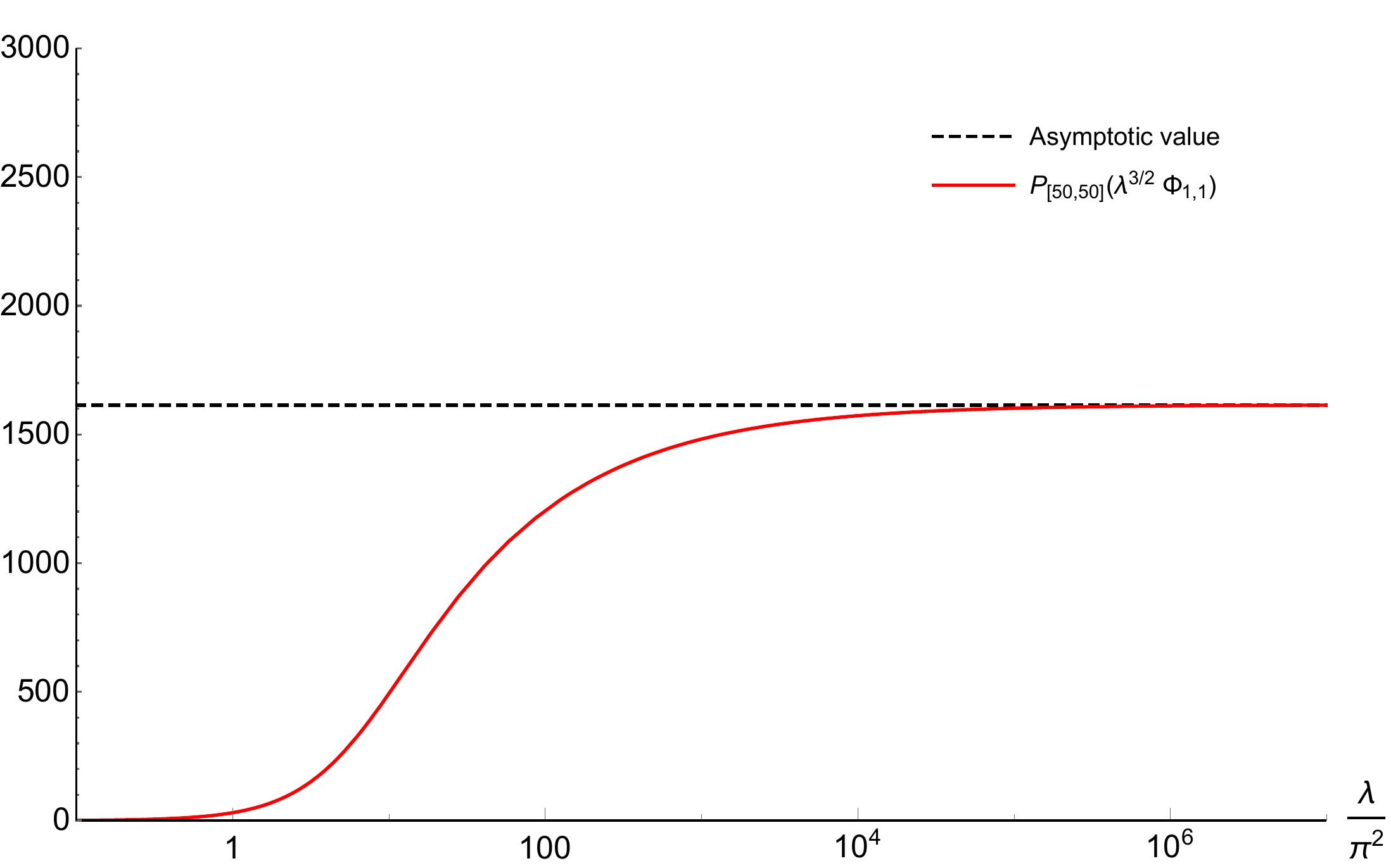}
    \caption{The red curve represents a diagonal conformal Pad\'e approximant of order 50 of the perturbative expansion of $\lambda^{\frac{3}{2}}\,\Phi_{1,1}$. The dashed black line represents its asymptotic constant value.}
    \label{fig:1}
\end{figure}

Using this analogy, on a very heuristic level we are led to expect that $\Phi_{q,r}$ is $O(\lambda^{-\frac{3}{2}})$ when $\lambda\to\infty$. In absence of an analytic proof, we have tested this expectation numerically, proceeding as follows. First, after expressing $\mathsf{D}^{\mathrm{odd}}$ in terms of the matrix $\mathsf{X}^{\mathrm{odd}}$ as in (\ref{Dnm}) and using the integral representation (\ref{Xij}) in terms of Bessel functions, we have generated very long expansions for $\Phi_{q,r}$ up to order $\lambda^{100}$, which represents a good compromise between the need to obtain sufficiently precise numerical results and the related computational cost. Then, we have resummed these expansions using a diagonal conformal Pad\'e approximant of order 50 which can be safely evaluated for values of $\lambda$ up to $\sim 10^7$ (see for example \cite{Beccaria:2021vuc,Beccaria:2021hvt} and references therein for details on this numerical method). In this way we managed to numerically check that the combinations
\begin{align}
    \lambda^{\frac{3}{2}}\,\Phi_{q,r}
\end{align}
tend to constants for large values of $\lambda$. We have done this check for $q,r=1,\ldots,4$, but 
our results suggest that this is a general behavior. As an illustrative example of our findings, we show in Fig.~\ref{fig:1} the plot of $\lambda^{\frac{3}{2}}\,\Phi_{1,1}$.

Since all terms in (\ref{DMD2}) are $O(\lambda^{-\frac{3}{2}})$, we can conclude that
\begin{align}
\label{DMD3}
    \big(\mathsf{D}^{\mathrm{odd}} \mathsf{M}^{\mathrm{odd}} \mathsf{D}^{\mathrm{odd}}\big)_{q,r}
     ~\underset{\lambda \rightarrow \infty}{\sim}~ O(\lambda^{-\frac{3}{2}})~,
\end{align}
as reported in (\ref{Co13}).

\printbibliography

\end{document}